\batchmode
\makeatletter
\def\input@path{{C:/Users/rlkam/Dropbox/Rushi/Research/Repo/tex/LaTeX/NCR/Rushi/AsymptoticSwitching/}}
\makeatother
\documentclass[english,draftcls, onecolumn]{IEEEtran}
\usepackage[T1]{fontenc}
\usepackage[latin9]{inputenc}
\usepackage{color}
\usepackage{units}
\usepackage{mathrsfs}
\usepackage{mathtools}
\usepackage{amsmath}
\usepackage{amsthm}
\usepackage{amssymb}
\usepackage{graphicx}
\usepackage{esint}
\usepackage{xargs}[2008/03/08]

\makeatletter
\theoremstyle{definition}
\newtheorem{assumption}{Assumption}
  \theoremstyle{plain}
  \newtheorem{prop}{\protect\propositionname}
  \theoremstyle{definition}
  \newtheorem{example}{\protect\examplename}
  \theoremstyle{definition}
  \newtheorem{defn}{\protect\definitionname}
  \theoremstyle{plain}
  \newtheorem{thm}{\protect\theoremname}
  \theoremstyle{plain}
  \newtheorem{cor}{\protect\corollaryname}
  \theoremstyle{remark}
  \newtheorem{rem}{\protect\remarkname}
  \theoremstyle{plain}
  \newtheorem{lem}{\protect\lemmaname}

\usepackage{cite}
\usepackage{tensor}
\newcommand\xqed[1]{%
  \leavevmode\unskip\penalty9999 \hbox{}\nobreak\hfill
  \quad\hbox{#1}}
\newcommand\defnEnd{\xqed{$\triangle$}}

\allowdisplaybreaks

\makeatother

\usepackage{babel}
\providecommand{\corollaryname}{Corollary}
\providecommand{\definitionname}{Definition}
\providecommand{\examplename}{Example}
\providecommand{\lemmaname}{Lemma}
\providecommand{\propositionname}{Proposition}
\providecommand{\remarkname}{Remark}
\providecommand{\theoremname}{Theorem}

\begin{document}

\title{Invariance-like Results for Nonautonomous Switched Systems\thanks{Rushikesh Kamalapurkar is with the School of Mechanical and Aerospace
Engineering, Oklahoma State University, Stillwater, OK, USA. Email:
rushikesh.kamalapurkar@okstate.edu.}\thanks{Joel A. Rosenfeld, Anup Parikh and Warren E. Dixon are with the Department
of Mechanical and Aerospace Engineering, University of Florida, Gainesville,
FL, USA. Email: \{joelar, anuppari, wdixon\}@ufl.edu.}\thanks{Andrew R. Teel is with the Department of Electrical and Computer Engineering,
University of California, Santa Barbara, CA, USA. Email: teel@ece.ucsb.edu}\thanks{This research is supported in part by NSF award numbers 1509516 and
1508757, ONR grant number N00014-13-1-0151, AFOSR Award Number FA9550-15-1-0155,
and a contract with the AFRL, Munitions Directorate at Eglin AFB.
Any opinions, findings and conclusions or recommendations expressed
in this material are those of the authors and do not necessarily reflect
the views of the sponsoring agency.}}

\author{Rushikesh Kamalapurkar, Joel A. Rosenfeld, Anup Parikh, Andrew R.
Teel, Warren E. Dixon}
\maketitle
\begin{abstract}
This paper generalizes the LaSalle-Yoshizawa Theorem to switched nonsmooth
systems. Filippov and Krasovskii regularizations of a switched system
are shown to be contained within the convex hull of the Filippov and
Krasovskii regularizations of the subsystems, respectively. A candidate
common Lyapunov function that has a negative semidefinite derivative
along the trajectories of the subsystems is shown to be sufficient
to establish LaSalle-Yoshizawa results for the switched system. Results
for regular and non-regular candidate Lyapunov functions are developed
via appropriate generalization of the notion of a time derivative.
The developed generalization is motivated by adaptive control of switched
systems where the derivative of the candidate Lyapunov function is
typically negative semidefinite.
\end{abstract}


\global\long\def\d{\textnormal{d}}

\global\long\def\sgn{\operatorname{sgn}}

\global\long\def\proj{\operatornamewithlimits{proj}}

\global\long\def\argmin{\operatornamewithlimits{arg\:min}}

\global\long\def\R{\mathbb{R}}

\global\long\def\tr{\operatorname{tr}}

\global\long\def\Sgn{\operatorname{SGN}}

\global\long\def\k{\operatorname{K}}

\global\long\def\co{\operatorname{co}}

\global\long\def\ae{\operatorname{a.e.}}

\global\long\def\b{\operatorname{B}}

\global\long\def\e{\operatorname{e}}

\global\long\def\n{\mathbb{N}}

\newcommandx\fid[5][usedefault, addprefix=\global, 1=, 2=, 3=, 4=, 5=]{\tensor*[_{#3}^{#2}]{#1}{_{#5}^{#4}}}
\textcolor{blue}{\global\long\def\d{\textnormal{d}}

\global\long\def\i{\textnormal{i}}

\global\long\def\sgn{\operatorname{sgn}}

\global\long\def\proj{\operatornamewithlimits{proj}}

\global\long\def\argmin{\operatornamewithlimits{arg\:min}}

\global\long\def\argmax{\operatornamewithlimits{arg\:max}}

\global\long\def\R{\mathbb{R}}

\global\long\def\tr{\operatorname{tr}}

\global\long\def\Sgn{\operatorname{SGN}}

\global\long\def\k{\operatorname{K}}

\global\long\def\co{\operatorname{co}}

\global\long\def\ae{\operatorname{a.e.}}

\global\long\def\b{\operatorname{B}}

\global\long\def\n{\mathbb{N}}

\global\long\def\diag{\operatorname{diag}}

\global\long\def\rank{\operatorname{rank}}

\global\long\def\ind{\operatorname{\mathbf{1}}}

\renewcommandx\fid[5][usedefault, addprefix=\global, 1=, 2=, 3=, 4=, 5=]{\tensor*[_{#3}^{#2}]{#1}{_{#5}^{#4}}}

\global\long\def\g{\operatorname{g}}

\global\long\def\eval#1{\left.#1\right|}

\global\long\def\lab#1#2{\underset{#2}{\underbrace{#1}}}

\global\long\def\zero{\operatorname{0}}

\global\long\def\e{\textnormal{e}}

\global\long\def\vecop{\operatorname{vec}}

\global\long\def\id{\operatorname{I}}

\global\long\def\re{\operatorname{re}}

\global\long\def\im{\operatorname{im}}

\global\long\def\lap#1{\mathscr{L}\left\{  #1\right\}  }

\global\long\def\dist{\operatorname{dist}}

\global\long\def\unit#1#2{\SI[per-mode=symbol]{#1}{#2}}

\global\long\def\ropen#1{\left[#1\right) }

\global\long\def\lopen#1{\left(#1\right]}

\global\long\def\lip{\operatorname{Lip}}

\global\long\def\cl{\operatornamewithlimits{cl}}

}

\section{\label{sec:Introduction}Introduction}

The focus of this paper is Lyapunov-based stability analysis of switched
nonautonomous systems that admit non-strict candidate Lyapunov functions
(cLfs) (i.e., cLfs with time derivatives bounded by a negative semidefinite
function of the state). The theoretical development is motivated by
the application of adaptive control methods to systems where either
the control design or the dynamics dictate the need for a switched
systems analysis. For example, neuromuscular electrical stimulation
applications such as \cite{Downey.Bellman.ea2015,Bellman.Cheng.ea2016,Rouse.Parikh.ea2016,Downey.Cheng.ea2017}
involve switching between different muscle groups during different
phases of operation to reduce fatigue \cite{Downey.Bellman.ea2015,Downey.Cheng.ea2017},
to compensate for changing muscle geometry \cite{Rouse.Parikh.ea2016},
or to perform functional tasks that require multi-limb coordination
\cite{Bellman.Cheng.ea2016}. Such applications stand to benefit from
adaptive methods where the controller adapts to the uncertain dynamics
without strictly relying on high gain or high frequency feedback
often associated with robust control methods that can lead to overstimulation.

Switched dynamics are inherent in a variety of modern adaptation strategies.
For example, in sparse neural networks \cite{Nivison.Khargonekar2017},
the use of different approximation architectures for different regions
of the state-space introduce switching via the feedforward part of
the controller. In adaptive gain scheduling methods \cite{Rugh.Shamma2000},
switching is introduced due to changing feedback gains. Switching
is also utilized as a tool to improve transient response of adaptive
controllers by selecting between multiple estimated models of stable
plants (see, e.g.,\cite{Narendra.Balakrishnan1994,Morse1996,Morse1997,Narendra.Balakrishnan1997,Hespanha.Liberzon.ea1999,Hespanha.Liberzon.ea2001,Anderson.Brinsmead.ea2001,Hespanha.Liberzon.ea2003,Hespanha.Liberzon.ea2003a,Vu.Chatterjee.ea2007}).
In addition, switched systems theory can be utilized to extend the
scope of existing adaptive solutions to more complex circumstances
that involve switched dynamics. 

Lyapunov-based stability analysis of switched nonautonomous adaptive
systems is challenging because adaptive update laws typically result
in non-strict Lyapunov functions for the individual subsystems. For
each subsystem, convergence of the error signal to the origin is typically
established using Barb{\u{a}}lat's lemma (e.g., \cite[Lemma 8.2]{Khalil2002}).\textit{\textcolor{blue}{{}
}}\textcolor{blue}{In traditional methods that utilize multiple Lyapunov
functions (e.g., \cite[Theorem 3.2]{Liberzon2003}) the class of admissible
switching signals is restricted based on the rate of decay of the
cLf (cf.\cite[Eq. 3.10]{Liberzon2003}). Since Barb{\u{a}}lat's lemma
provides no information about the rate of decay of the cLf, it alone
is insufficient to establish stability of a switched system using
multiple Lyapunov functions.} Approaches based on common cLfs have
been developed for systems with negative definite Lyapunov derivatives;
however, common cLf-based approaches do not trivially extend to systems
with non-strict Lyapunov functions (cf.\cite{Poveda.Teel2017,Mancilla-Aguilar.Garcia.ea2005,Mancilla-Aguilar.Garcia.ea2005a}
and \cite[Example 2.1]{Liberzon2003}).

Because of complications resulting from a negative semidefinite Lyapunov
derivative, few results are available in literature that examine adaptive
control of uncertain nonlinear switched systems. An adaptive controller
for switched nonlinear systems is developed in \cite{Long.Wang.ea2015}
using a generalization of Barb{\u{a}}lat's lemma from \cite{Jiang2009}.
The controller is shown to asymptotically stabilize a switched system
with parametric uncertainties in the subsystems. Multiple Lyapunov
functions are utilized to analyze the stability of the switched system.
However, the generalized Barb{\u{a}}lat's Lemma in \cite{Jiang2009}
requires a minimum dwell time, and in general, minimum dwell time
cannot be guaranteed when the switching is state-dependent.

Results such as \cite{Ryan1998,Logemann.Ryan2004,Bacciotti.Mazzi2005,Hui.Haddad.ea2009}
extend the Barbashin-Krasovskii-LaSalle invariance principle to discontinuous
systems. However, these results are for autonomous systems, whereas
the development in this paper is focused on nonautonomous systems.
An extension of the LaSalle-Yoshizawa Theorem to nonsmooth nonautonomous
systems is provided in \cite[Theorem 2.5]{Haddad2006}; however, the
result requires the cLf to be continuously differentiable, whereas
the approach developed here uses a more general framework that utilizes
locally Lipschitz-continuous cLfs.

This paper generalizes the LaSalle-Yoshizawa Theorem (see, e.g.,\cite{Yoshizawa1963}
and \cite[Theorem 8.4]{Khalil2002}) and its nonsmooth extensions
in results such as \cite[Theorem 2.5]{Haddad2006}, and \cite{Fischer.Kamalapurkar.ea2013}
to switched nonsmooth systems and nonregular Lyapunov functions. A
non-strict common Lyapunov function (i.e., a common cLf with a negative
semidefinite derivative) is used to establish boundedness of the system
state and convergence of a positive semidefinite function of the system
state to zero under arbitrary switching between nonsmooth nonlinear
systems.

\textcolor{blue}{The paper is organized as follows. Notation is defined
in Section \ref{sec:Notation}. Section \ref{sec:ProblemDescription}
defines the class of systems considered along with the objectives.
Sections \ref{sec:Switching-and-regularization}-\ref{sec:Main-Result}
are dedicated to the development of the main results of the paper.
Section \ref{sec:Comments-on-the} provides a discussion on the merits
of the generalized time derivatives defined in Section \ref{sec:Semidefinite-Lyapunov-functions}.
Section \ref{sec:Example} presents illustrative examples, and Section
\ref{sec:Conclusion} provides concluding remarks. The appendix includes
supplementary proofs.}

\section{\label{sec:Notation}Notation}

\textcolor{blue}{The $n-$dimensional Euclidean space is denoted by
$\R^{n}$ and $\mu$ denotes the Lebesgue measure on $\R^{n}$. Elements
of $\R^{n}$ are interpreted as column vectors and $\left(\cdot\right)^{T}$
denotes the vector transpose operator. The set of positive integers
excluding 0 is denoted by $\n$. For $a\in\R,$ the notation $\R_{\geq a}$
denotes the interval $\left[a,\infty\right)$ and the notation $\R_{>a}$
denotes the interval $\left(a,\infty\right)$. For a relation $\left(\cdot\right)$,
the notation $\overset{\ae}{\left(\cdot\right)}$ implies that the
relation holds for almost all $t\in\mathcal{I}$, for some interval
$\mathcal{I}$. Unless otherwise specified, an interval $\mathcal{I}$
is assumed to be right open, of nonzero length, and $t_{0}\coloneqq\min\mathcal{I}$.
The notation $F:A\rightrightarrows B$ is used to denote a set-valued
map from $A$ to the subsets of $B$. The notations $\co A$ and $\overline{\co}A$
are used to denote the convex hull and the closed convex hull of the
set $A$, respectively. If $a\in\R^{m}$ and $b\in\R^{n}$ then the
notation $\left[a;b\right]$ denotes the concatenated vector $\begin{bmatrix}a\\
b
\end{bmatrix}\in\R^{m+n}$. For $A\subseteq\R^{m}$, $B\subseteq\R^{n}$ the notations $\begin{bmatrix}A\\
B
\end{bmatrix}$ and $A\times B$ are interchangeably used to denotes the set $\left\{ \left[a;b\right]\mid a\in A,b\in B\right\} $.
The notations $\overline{\b}\left(x,r\right)$ and $\b\left(x,r\right)$
for $x\in\R^{n}$ and $r>0$ are used to denote the sets $\left\{ y\in\R^{n}\mid\left\Vert x-y\right\Vert \leq r\right\} $
and $\left\{ y\in\R^{n}\mid\left\Vert x-y\right\Vert <r\right\} $,
respectively. The notation $\left|\left(\cdot\right)\right|$ denotes
the absolute value if $\left(\cdot\right)\in\R$ and the cardinality
if $\left(\cdot\right)$ is a set. The notation $\mathcal{L}_{\infty}\left(A,B\right)$
denotes essentially bounded functions from $A$ to $B$.}

\section{\label{sec:ProblemDescription}Switched systems and differential
inclusions}

Consider a switched system of the form
\begin{equation}
\dot{x}=f_{\rho\left(x,t\right)}\left(x,t\right),\label{eq:Dyn}
\end{equation}
where $\rho:\R^{n}\times\R_{\geq t_{0}}\to\mathcal{N}^{o}$ denotes
a state-dependent switching signal, $\mathcal{N}^{o}\subseteq\mathbb{N}$
is the set of all possible switching indices, and $x\in\R^{n}$ denotes
the system state. The collection $\left\{ f_{\sigma}:\R^{n}\times\R_{\geq t_{0}}\to\R^{n}\right\} _{\sigma\in\mathcal{N}^{o}}$
is assumed to be locally bounded, uniformly in $\sigma$ and $t$,\footnote{A collection of functions $\left\{ f_{\sigma}:\R^{n}\times\R_{\geq t_{0}}\to\R^{n}\mid\sigma\in\mathcal{N}^{o}\right\} $
is locally bounded, uniformly in $t$ and $\sigma$, if for every
compact $K\subset\R^{n}$, there exists $M>0$ such that $\left\Vert f_{\sigma}\left(x,t\right)\right\Vert _{2}\leq M,\:\forall\left(x,t\right)\in K\times\R_{\geq t_{0}}$
and $\forall\sigma\in\mathcal{N}^{o}$.} and the functions $t\mapsto f_{\sigma}\left(x,t\right)$ and $t\mapsto\rho\left(x,t\right)$
are assumed to be Lebesgue measurable $\forall x\in\R^{n}$ and $\forall\sigma\in\mathcal{N}^{o}$.\textcolor{blue}{{} }

Let $f:\R^{n}\times\R_{\geq t_{0}}\to\R^{n}$ denote the function
$f\left(x,t\right)\coloneqq f_{\rho\left(x,t\right)}\left(x,t\right).$
Since the collection $\left\{ f_{\sigma}\right\} _{\sigma\in\mathcal{N}^{o}}$
locally bounded, uniformly in $\sigma$ and $t$, the function $f$
is locally bounded, uniformly in $t$. To establish measurability
of $f$, consider the representation $f\left(x,t\right)=\sum_{\sigma\in\mathcal{N}^{o}}I_{\sigma}\left(\rho\left(x,t\right)\right)f_{\sigma}\left(x,t\right)$,
where 
\[
I_{\sigma}\left(i\right)\coloneqq\begin{cases}
1, & i=\sigma,\\
0, & i\neq\sigma.
\end{cases}
\]
Since $I_{\sigma}:\mathbb{N}\to\R$ is continuous $\forall\sigma\in\mathcal{N}^{o}$,
$t\mapsto I_{\sigma}\left(\rho\left(x,t\right)\right)$ is Lebesgue
measurable $\forall\left(\sigma,x\right)\in\mathcal{N}^{o}\times\R^{n}$.
Lebesgue measurability of $t\mapsto f\left(x,t\right)$, $\forall x\in\R^{n}$
then follows from that of $t\mapsto f_{\sigma}\left(x,t\right)$,
$\forall\left(\sigma,x\right)\in\mathcal{N}^{o}\times\R^{n}$. 

The main objective of this paper is to establish asymptotic properties
of the generalized solutions to the system 
\begin{equation}
\dot{x}=f\left(x,t\right),\label{eq:Switched System}
\end{equation}
using asymptotic properties of the generalized solutions to the individual
subsystems 
\begin{equation}
\dot{x}=f_{\sigma}\left(x,t\right).\label{eq:Subsystems}
\end{equation}
\textcolor{blue}{The advantage of the strategy pursued in this paper,
as opposed to directly analyzing (\ref{eq:Switched System}), is that
the analysis can be made independent of the switching function. That
is, when established through the subsystems in (\ref{eq:Subsystems}),
the stability properties of (\ref{eq:Switched System}) are invariant
with respective to the switching function over a wide range of switching
functions. On the other hand, a direct analysis of (\ref{eq:Switched System})
is valid only for the specific $\rho$ used to construct (\ref{eq:Switched System}).}

\textcolor{blue}{For a measurable function $g:\R^{n}\times\R_{\geq t_{0}}\to\R$
}the Filippov regularization is defined as\cite[p. 85]{Filippov1988}
\begin{equation}
\k_{\mathbb{F}}\left[g\right]\left(x,t\right)\coloneqq\bigcap_{\delta>0}\bigcap_{\mu\left(N\right)=0}\overline{\co}\left\{ g\left(y,t\right)\mid y\in\b\left(x,\delta\right)\setminus N\right\} ,\label{eq:Filippov}
\end{equation}
and the Krasovskii regularization is defined as\cite[p. 17]{Krasovskii.Subbotin1988}
\begin{equation}
\k_{\mathbb{K}}\left[g\right]\left(x,t\right)\coloneqq\bigcap_{\delta>0}\overline{\co}\left\{ g\left(y,t\right)\mid y\in\b\left(x,\delta\right)\right\} .\label{eq:Krasovskii}
\end{equation}
In the following, generalized solutions of the systems in (\ref{eq:Switched System})
and (\ref{eq:Subsystems}), defined using Filippov and Krasovskii
regularization are analyzed. When a Filippov regularization is considered,
the local boundedness requirement on the map $x\mapsto f_{\sigma}\left(x,t\right)$
is relaxed to essential local boundedness and a stronger measurability
requirement is imposed so that $\left(x,t\right)\mapsto f_{\sigma}\left(x,t\right)$
and $\left(x,t\right)\mapsto\rho\left(x,t\right)$ are Lebesgue measurable
$\forall\sigma\in\mathcal{N}^{o}$. 

\textcolor{blue}{To achieve the stated objective, the differential
inclusion that results from regularization of the overall switched
system is proven to be contained within the convex combination of
the differential inclusions that result from regularization of the
subsystems under mild assumptions on the switching signal (Proposition
\ref{prop:The-set-valued-maps}, Section \ref{sec:Switching-and-regularization}).
To facilitate the discussion that follows, the existence of a non-strict
Lyapunov function is shown to be sufficient to infer certain asymptotic
properties of solutions to differential inclusions (Theorem \ref{thm:GLYT},
Section \ref{sec:Semidefinite-Lyapunov-functions}). It is then established
that a common non-strict Lyapunov function for the differential inclusions
that result from regularization of the individual subsystems is also
a non-strict Lyapunov function for the differential inclusion that
results from regularization of the switched system (Proposition \ref{prop:Final},
Section \ref{sec:Main-Result}). The main result of the paper then
follows, i.e., conclusions about asymptotic properties of generalized
solutions to (\ref{eq:Switched System}) can be drawn from the asymptotic
properties of the generalized solutions to (\ref{eq:Subsystems})
(Theorem \ref{thm:FinalResult}).}

\textcolor{blue}{The following section develops the aforementioned
relationship between the differential inclusions resulting from regularization
of the subsystems and the switched system.}

\section{\label{sec:Switching-and-regularization}Switching and regularization}

\textcolor{blue}{Let $\dot{x}\in\mathbb{F}\left(x,t\right)\coloneqq\k_{\mathbb{F}}\left[f\right]\left(x,t\right)$
and $\dot{x}\in\mathbb{F}_{\sigma}\left(x,t\right)\coloneqq\k_{\mathbb{F}}\left[f_{\sigma}\right]\left(x,t\right)$
be Filippov regularizations and $\dot{x}\in\mathbb{K}\left(x,t\right)\coloneqq\k_{\mathbb{K}}\left[f\right]\left(x,t\right)$
and $\dot{x}\in\mathbb{K}_{\sigma}\left(x,t\right)\coloneqq\k_{\mathbb{K}}\left[f_{\sigma}\right]\left(x,t\right)$
be Krasovskii regularizations of (\ref{eq:Switched System}) and (\ref{eq:Subsystems}),
respectively. The following assumption imposes a mild restriction
on the switching function $\rho$ to establish a relationship between
$\mathbb{F}$, $\left\{ \mathbb{F}_{\sigma}\right\} $, $\mathbb{K}$,
and $\left\{ \mathbb{K}_{\sigma}\right\} $. }
\begin{assumption}
\textcolor{blue}{\label{assu:locally-finite-switches}For each $\left(x,t\right)\in\mathbb{R}^{n}\times\mathbb{R}_{\geq t_{0}}$,
there exists $\delta^{*}>0$ such that $\left|\rho\left(\textnormal{B}\left(x,\delta^{*}\right),t\right)\right|<\infty$.}\defnEnd
\end{assumption}
\textcolor{blue}{That is, that $\rho$ is locally bounded in $x$
for each $t$. Roughly speaking, Assumption \ref{assu:locally-finite-switches}
restricts infinitely many subsystems from being active in a small
neighborhood of the state space. It does not restrict Zeno behavior
and arbitrary time-dependent switching, and as such, is not restrictive.
For further insight into why Assumption \ref{assu:locally-finite-switches}
is invoked, see Example \ref{exa:Let--and}. The following proposition
states that under general conditions, the set-valued maps $\mathbb{F}$
and $\mathbb{K}$ are contained, pointwise, within the convex combination
of the collections $\left\{ \mathbb{F}_{\sigma}\right\} $ and $\left\{ \mathbb{K}_{\sigma}\right\} $,
respectively.}
\begin{prop}
\label{prop:The-set-valued-maps}\textcolor{blue}{Provided $\rho:\R^{n}\times\R_{\geq t_{0}}\to\mathcal{N}^{o}$
satisfies Assumption \ref{assu:locally-finite-switches},} there exists
a set $E\subseteq\R_{\geq t_{0}}$ with $\mu\left(\R_{\geq t_{0}}\setminus E\right)=0$
such that the set-valued maps $\mathbb{F}$, $\mathbb{F}_{\sigma}$,
$\mathbb{K}$, and $\mathbb{K}_{\sigma}$ satisfy
\begin{align}
\mathbb{K}\left(x,t\right) & \subseteq\overline{\co}\bigcup_{\sigma\in\mathcal{N}^{o}}\mathbb{K}_{\sigma}\left(x,t\right),\quad\forall\left(x,t\right)\in\R^{n}\times\R_{\geq t_{0}},\label{eq:Convex Hull Condition Krasovskii}\\
\mathbb{F}\left(x,t\right) & \subseteq\overline{\co}\bigcup_{\sigma\in\mathcal{N}^{o}}\mathbb{F}_{\sigma}\left(x,t\right),\quad\forall\left(x,t\right)\in\R^{n}\times E.\label{eq:Convex Hull Almost Everywhere}
\end{align}
Under the additional assumption that $\forall\sigma\in\mathcal{N}^{o}$,
there exist countable collections of measure-zero sets $\left\{ N_{\sigma i}\right\} _{i\in\mathbb{N}}$,
\textcolor{blue}{and a constant $\overline{\delta}>0$} such that
$\forall\left(x,t\right)\in\R^{n}\times\R_{\geq t_{0}}$ and \textcolor{blue}{$\forall\delta\in\left(0,\overline{\delta}\right]$},\footnote{\textcolor{blue}{The condition in (\ref{eq:FilippovCountableAssumption})
is satisfied by most discontinuous dynamical systems encountered in
practice. For example, discontinuities resulting from sliding mode
controllers, piecewise continuous reference signals, etc., satisfy
(\ref{eq:FilippovCountableAssumption}). Hence, (\ref{eq:FilippovCountableAssumption})
is not restrictive in practice.}}
\begin{multline}
\bigcap_{\mu\left(N\right)=0}\overline{\co}\left\{ f_{\sigma}\left(y,t\right)\mid y\in\b\left(x,\delta\right)\setminus N\right\} =\\
\bigcap_{i\in\mathbb{N}}\overline{\co}\left\{ f_{\sigma}\left(y,t\right)\mid y\in\b\left(x,\delta\right)\setminus N_{\sigma i}\right\} ,\label{eq:FilippovCountableAssumption}
\end{multline}
the inclusion in (\ref{eq:Convex Hull Almost Everywhere}) holds with
$E=\R_{\geq t_{0}}$.
\end{prop}
\begin{IEEEproof}[Proof for Krasovskii regularization]
\textcolor{blue}{Fix $\left(x,t\right)\in\R^{n}\times\R_{\geq t_{0}}$,
select $\delta^{*}>0$ such that $\left|\rho\left(\textnormal{B}\left(x,\delta^{*}\right),t\right)\right|<\infty$,}\footnote{\textcolor{blue}{Existence of such a $\delta^{*}$ is guaranteed by
Assumption \ref{assu:locally-finite-switches}.}}\textcolor{blue}{{} and let $\mathcal{N}\coloneqq\rho\left(\textnormal{B}\left(x,\delta^{*}\right),t\right)$.}
\textcolor{blue}{Observe that the containment in (\ref{eq:Convex Hull Condition Krasovskii})
is straightforward if the union over $\sigma$ is placed inside the
convex closure operation. That is,
\begin{multline}
\bigcap_{\delta>0}\overline{\co}\bigl\{ f_{\rho\left(y,t\right)}\left(y,t\right)\mid y\in\b\left(x,\delta\right)\bigr\}\subseteq\\
\bigcap_{\delta>0}\overline{\co}\bigcup_{\sigma\in\mathcal{N}}\bigl\{ f_{\sigma}\left(y,t\right)\mid y\in\b\left(x,\delta\right)\bigr\}.\label{eq:fist}
\end{multline}
The rest of the proof shows that the right hand side (RHS) of (\ref{eq:fist})
is contained within the RHS of (\ref{eq:Convex Hull Condition Krasovskii})
in two steps. The first step is to show that
\begin{multline}
\bigcap_{\delta>0}\overline{\co}\bigcup_{\sigma\in\mathcal{N}}\bigl\{ f_{\sigma}\left(y,t\right)\mid y\in\b\left(x,\delta\right)\bigr\}\subseteq\\
\bigcap_{\delta>0}\co\bigcup_{\sigma\in\mathcal{N}}\overline{\co}\bigl\{ f_{\sigma}\left(y,t\right)\mid y\in\b\left(x,\delta\right)\bigr\}.\label{eq:steptwo}
\end{multline}
The second step is to show that 
\begin{multline}
\bigcap_{\delta>0}\co\bigcup_{\sigma\in\mathcal{N}}\overline{\co}\bigl\{ f_{\sigma}\left(y,t\right)\mid y\in\b\left(x,\delta\right)\bigr\}\subseteq\\
\co\bigcup_{\sigma\in\mathcal{N}}\bigcap_{\delta>0}\overline{\co}\left\{ f_{\sigma}\left(y,t\right)\mid y\in\b\left(x,\delta\right)\right\} .\label{eq:stepthree}
\end{multline}
The result in (\ref{eq:Convex Hull Condition Krasovskii}) then follows
from (\ref{eq:fist}), (\ref{eq:steptwo}), and (\ref{eq:stepthree}).}

\textcolor{blue}{To prove (\ref{eq:steptwo}),} fix $\delta\in\left(0,\delta^{*}\right]$
and let $z\in\overline{\co}\bigcup_{\sigma\in\mathcal{N}}\bigl\{ f_{\sigma}\left(y,t\right)\mid y\in\b\left(x,\delta\right)\bigr\}$.
There exists a sequence $\left\{ z_{i}\right\} _{i\in\mathbb{N}}\in\R^{n}$
such that $z_{i}\in\co\bigcup_{\sigma\in\mathcal{N}}\bigl\{ f_{\sigma}\left(y,t\right)\mid y\in\b\left(x,\delta\right)\bigr\}$,
$\forall i\in\mathbb{N}$, and $\lim_{i\to\infty}z_{i}=z$. Furthermore,
by Carath\'{e}odory's Theorem \cite[p. 103]{Danzer.Gruenbaum.ea1963},
there exists collection of $m\le n+1$ points $\left\{ z_{i1},\cdots,z_{im}\right\} \subset\R^{n}$,
positive real numbers $\left\{ a_{i1},\cdots,a_{im}\right\} $ for
which $\sum_{j=1}^{m}a_{ij}=1$, and integers $\left\{ \sigma_{i1},\cdots,\sigma_{im}\right\} \in\mathcal{N}$,
such that $z_{ij}\in\bigl\{ f_{\sigma_{j}}\left(y,t\right)\mid y\in\b\left(x,\delta\right)\bigr\}$
and $z_{i}=\sum_{j=1}^{m}a_{ij}z_{ij}$. Hence, $z=\lim_{i\to\infty}\sum_{j=1}^{m}a_{ij}z_{ij}$,
that is, $z=\lim_{i\to\infty}Z_{i}A_{i}$, where $A_{i}=\begin{bmatrix}a_{i1} & ;\cdots; & a_{im}\end{bmatrix}$
and $Z_{i}=\begin{bmatrix}z_{i1}; & \cdots; & z_{im}\end{bmatrix}^{T}$.

Since the coefficients $a_{ij}\geq0$ are bounded, the sequence $\left\{ A_{i}\right\} _{i\in\mathbb{N}}$
is a bounded sequence. Hence, there exists a subsequence $\left\{ A_{i_{k}}\right\} _{k\in\mathbb{N}}$
such that $\lim_{k\to\infty}A_{i_{k}}=A,$ for some $A=\begin{bmatrix}a_{1} & ;\cdots; & a_{m}\end{bmatrix}$.
Since the function $A_{i}\mapsto\sum_{j=1}^{m}a_{ij}$ is continuous,
$\sum_{j=1}^{m}a_{j}=1$. Since the set $\bigcup_{\sigma\in\mathcal{N}}\left\{ f_{\sigma}\left(y,t\right)\mid y\in\b\left(x,\delta\right)\right\} $
is bounded, the sequence $\left\{ Z_{i_{k}}\right\} _{k\in\mathbb{N}}$
is bounded. Hence there exists a subsequence $\left\{ Z_{i_{k_{l}}}\right\} _{l\in\mathbb{N}}$
such that $\lim_{l\to\infty}Z_{i_{k_{l}}}=Z$, element-wise, for some
$Z=\begin{bmatrix}z_{1} & ;\cdots; & z_{m}\end{bmatrix}^{T}$. Hence,
$z=\lim_{l\to\infty}Z_{i_{k_{l}}}A_{i_{k_{l}}}=ZA$, where the columns
$z_{j}$ of the matrix $Z$ are the limits $\lim_{l\to\infty}z_{i_{k_{l}}j}$.
Hence, $z_{j}\in\overline{\co}\bigl\{ f_{\sigma_{j}}\left(y,t\right)\mid y\in\b\left(x,\delta\right)\bigr\}$.
Therefore, the point $z$ is a convex combination of points from $\overline{\co}\bigl\{ f_{\sigma_{j}}\left(y,t\right)\mid y\in\b\left(x,\delta\right)\bigr\}$.
That is, $z\in\co\bigcup_{\sigma\in\mathcal{N}}\overline{\co}\left\{ f_{\sigma}\left(y,t\right)\mid y\in\b\left(x,\delta\right)\right\} $\textcolor{blue}{{}
$\forall\delta\in\left(0,\delta^{*}\right]$; which proves} (\ref{eq:steptwo}).

\textcolor{blue}{To establish (\ref{eq:stepthree}),} let $z\in\bigcap_{\delta>0}\co\bigcup_{\sigma\in\mathcal{N}}\overline{\co}\bigl\{ f_{\sigma}\left(y,t\right)\mid y\in\b\left(x,\delta\right)\bigr\}$.
Note that if $0<\delta_{1}\leq\delta_{2}$, then
\begin{multline*}
\co\bigcup_{\sigma\in\mathcal{N}}\overline{\co}\bigl\{ f_{\sigma}\left(y,t\right)\mid y\in\b\left(x,\delta_{1}\right)\bigr\}\subseteq\\
\co\bigcup_{\sigma\in\mathcal{N}}\overline{\co}\bigl\{ f_{\sigma}\left(y,t\right)\mid y\in\b\left(x,\delta_{2}\right)\bigr\}.
\end{multline*}
\textcolor{blue}{That is, if $z\in\co\bigcup_{\sigma\in\mathcal{N}}\overline{\co}\bigl\{ f_{\sigma}\left(y,t\right)\mid y\in\b\left(x,\delta_{1}\right)\bigr\}$
for some $0<\delta_{1}$, then $z\in\bigcap_{\delta>\delta_{1}}\co\bigcup_{\sigma\in\mathcal{N}}\overline{\co}\bigl\{ f_{\sigma}\left(y,t\right)\mid y\in\b\left(x,\delta\right)\bigr\}$.
Hence, $\forall k\in\n$, such that $k\geq\frac{1}{\delta^{*}}$,
there exist $\bigl\{ z_{k1},\cdots,z_{k\left|\mathcal{N}\right|}\bigr\}\subset\R^{n}$,
nonnegative real numbers $\bigl\{ a_{k1},\cdots,a_{k\left|\mathcal{N}\right|}\bigr\}$
for which $\sum_{j=1}^{\left|\mathcal{N}\right|}a_{kj}=1$, such that
$z_{kj}\in\bigcap_{\delta\geq\frac{1}{k}}\overline{\co}\bigl\{ f_{\sigma_{j}}\left(y,t\right)\mid y\in\b\left(x,\delta\right)\bigr\}$
and $z=\sum_{j=1}^{\left|\mathcal{N}\right|}a_{kj}z_{kj}$. That is,
$z=Z_{k}A_{k}$, where $A_{k}=\begin{bmatrix}a_{k1} & ;\cdots; & a_{k\left|\mathcal{N}\right|}\end{bmatrix}$
and $Z_{k}=\begin{bmatrix}z_{k1}; & \cdots; & z_{k\left|\mathcal{N}\right|}\end{bmatrix}^{T}$.
Since the sequences $\left\{ Z_{k}\right\} _{k\in\mathbb{N}}$ and
$\left\{ A_{k}\right\} _{k\in\mathbb{N}}$ are bounded, there exist
subsequences $\left\{ Z_{k_{l}}\right\} _{l\in\mathbb{N}}$ and $\left\{ A_{k_{l}}\right\} _{l\in\mathbb{N}}$
and vectors $Z\coloneqq\begin{bmatrix}z_{1} & ;\cdots; & z_{\left|\mathcal{N}\right|}\end{bmatrix}^{T}$
and $A\coloneqq\begin{bmatrix}a_{1} & ;\cdots; & a_{\left|\mathcal{N}\right|}\end{bmatrix}$
such that $A=\lim_{l\to\infty}A_{k_{l}}$, $\sum_{j=1}^{\left|\mathcal{N}\right|}a_{j}=1$,
and $Z=\lim_{l\to\infty}Z_{k_{l}}$. Since $z=Z_{k_{l}}A_{k_{l}},$
$\forall k_{l}\in\n$, it can be concluded that $z=ZA$.}

\textcolor{blue}{It is now claimed that $\forall j\in\left\{ 1,\cdots,\left|\mathcal{N}\right|\right\} ,$
$z_{j}\in\bigcap_{\delta>0}\overline{\co}\bigl\{ f_{\sigma_{j}}\left(y,t\right)\mid y\in\b\left(x,\delta\right)\bigr\}$.
To prove the claim by contradiction, assume that $\exists\delta^{*}>0$
such that $z_{j}\notin\overline{\co}\bigl\{ f_{\sigma_{j}}\left(y,t\right)\mid y\in\b\left(x,\delta^{*}\right)\bigr\}$.
Since
\begin{equation}
\overline{\co}\bigl\{\!f_{\sigma_{j}}\!\left(y,t\right)\!\mid\!y\!\in\!\b\left(x,\delta_{1}\right)\!\bigr\}\!\subseteq\!\overline{\co}\bigl\{\!f_{\sigma_{j}}\!\left(y,t\right)\!\mid\!y\!\in\!\b\left(x,\delta_{2}\right)\!\bigr\}\!,\label{eq:nested}
\end{equation}
$\forall\sigma_{j}\in\mathcal{N}$ and $\forall\delta_{1}\leq\delta_{2}$,
$z_{j}\notin\overline{\co}\bigl\{ f_{\sigma_{j}}\left(y,t\right)\mid y\in\b\left(x,\delta\right)\bigr\}$,
$\forall\delta\geq\delta^{*}$. That is, $\exists k_{l}^{*}\in\n$
such that $z_{j}\notin\bigcap_{\delta\geq\frac{1}{k_{l}}}\overline{\co}\bigl\{ f_{\sigma_{j}}\left(y,t\right)\mid y\in\b\left(x,\delta\right)\bigr\}$,
$\forall k_{l}\geq k_{l}^{*}$. From }(\ref{eq:nested})\textcolor{blue}{{}
and the fact that the sets $\bigcap_{\delta\geq\frac{1}{k_{l}}}\overline{\co}\bigl\{ f_{\sigma_{j}}\left(y,t\right)\mid y\in\b\left(x,\delta\right)\bigr\}$
are closed, it can be concluded that there exists $\epsilon>0$ such
that $\forall k_{l}\geq k_{l}^{*}$,
\begin{equation}
\b\left(z_{j},\epsilon\right)\notin\bigcap_{\delta\geq\frac{1}{k_{l}}}\overline{\co}\bigl\{ f_{\sigma_{j}}\left(y,t\right)\mid y\in\b\left(x,\delta\right)\bigr\}.\label{eq:contradict}
\end{equation}
Since $z_{k_{l}j}\in\bigcap_{\delta\geq\frac{1}{k_{l}}}\overline{\co}\bigl\{ f_{\sigma_{j}}\left(y,t\right)\mid y\in\b\left(x,\delta\right)\bigr\}$,
$\forall k_{l}\in\n$, }(\ref{eq:contradict})\textcolor{blue}{{} contradicts
$z_{j}=\lim_{l\to\infty}z_{k_{l}j}$.}

\textcolor{blue}{Hence, $z$ is a convex combination of points from
$\bigcap_{\delta>0}\overline{\co}\bigl\{ f_{\sigma_{j}}\left(y,t\right)\mid y\in\b\left(x,\delta\right)\bigr\}$.
That is, $z\in\co\bigcup_{\sigma\in\mathcal{N}}\bigcap_{\delta>0}\overline{\co}\left\{ f_{\sigma}\left(y,t\right)\mid y\in\b\left(x,\delta\right)\right\} $,
which proves (\ref{eq:stepthree}), and hence, (\ref{eq:Convex Hull Condition Krasovskii}).
The proof for Filippov regularization involves technical details related
to exclusion of measure-zero sets that are provided in the appendix.}
\end{IEEEproof}
\textcolor{blue}{The following example demonstrates that Assumption
\ref{assu:locally-finite-switches} is not vacuous.}\footnote{\textcolor{blue}{The authors are grateful to the anonymous reviewer
who suggested this example. }}
\begin{example}
\textcolor{blue}{\label{exa:Let--and}Let $\mathcal{N}^{o}=\mathbb{N}$
and for $\sigma\in\mathcal{N}^{o}$, let $f_{\sigma}$ be defined
as 
\[
f_{\sigma}\left(x\right)\coloneqq\begin{cases}
0 & \left|x\right|<\nicefrac{1}{2^{\sigma}}\\
1 & \left|x\right|\geq\nicefrac{1}{2^{\sigma}}
\end{cases},
\]
so that $\mathbb{K}_{\sigma}\left(0\right)=\mathbb{F}_{\sigma}\left(0\right)=\left\{ 0\right\} ,$
$\forall\sigma\in\mathcal{N}^{o}$. Let 
\[
\rho\left(x\right)=\begin{cases}
\sigma & x\in\left(-\frac{1}{2^{\sigma-1}},-\frac{1}{2^{\sigma}}\right]\cup\left[\frac{1}{2^{\sigma}},\frac{1}{2^{\sigma-1}}\right)\\
1 & \textnormal{otherwise}
\end{cases}.
\]
Clearly, $\rho$ violates Assumption \ref{assu:locally-finite-switches}
at $x=0$. In this case, $f\left(x\right)=\begin{cases}
1 & x\neq0\\
0 & x=0
\end{cases}.$ Therefore, $\mathbb{K}\left(0\right)=\left[0,1\right]$ and $\mathbb{F}\left(0\right)=\left\{ 1\right\} $,
and hence, Proposition \ref{prop:The-set-valued-maps} does not hold
without the switching restriction in Assumption \ref{assu:locally-finite-switches}.\defnEnd}
\end{example}
\textcolor{blue}{To facilitate the analysis of $\mathbb{F}$ and
$\mathbb{K}$ based on the analysis of the individual subsystems $\mathbb{F}_{\sigma}$
and $\mathbb{K}_{\sigma}$, respectively, a stability result for differential
inclusions that relies on non-strict Lyapunov functions is developed
in the following section. While the results developed in Section \ref{sec:Switching-and-regularization}
are specific to differential inclusions that arise from Filippov and
Krasovskii regularization of differential equations with discontinuous
right-hand sides, the results developed in the following sections
are more general in the sense that they apply to generic set-valued
maps not necessarily resulting from Filippov or Krasovskii regularization.}

\section{\label{sec:Semidefinite-Lyapunov-functions}Non-strict Lyapunov functions
for differential inclusions}

\textcolor{green}{}\textcolor{red}{Let $F:\mathcal{\R}^{n}\times\R_{\geq t_{0}}\rightrightarrows\mathcal{\R}^{n}$
be a set-valued map.} Consider a differential inclusion of the form
\begin{equation}
\dot{x}\in F\left(x,t\right).\label{eq:Inclusion}
\end{equation}
A locally absolutely continuous function $x:\mathcal{I}\to\mathcal{\R}^{n}$
is called a solution of (\ref{eq:Inclusion}) over the closed interval
$\mathcal{I}$ provided
\begin{equation}
\dot{x}\left(t\right)\in F\left(x\left(t\right),t\right),\label{eq:GenSol}
\end{equation}
for almost all $t\in\mathcal{I}$ \cite[p. 50]{Filippov1988}.\textcolor{red}{{}
To ensure existence of local solutions, the following restrictions
are placed on the map $F$.}
\begin{assumption}
\textcolor{red}{\label{ass:existence}For all $\left(x_{0},t_{0}\right)\in\R^{n}\times\R_{\geq0}$,
there exist $a,b\in\R_{>0}$ such that $F:\R^{n}\times\R_{\geq t_{0}}\rightrightarrows\R^{n}$
satisfies the hypothesis of \cite[p. 83, Theorem 5]{Filippov1988}.}
\end{assumption}
\textcolor{red}{Under Assumption \ref{ass:existence}, local solutions
of (\ref{eq:Inclusion}) exists starting from any $\left(t_{0},x_{0}\right)\in\R^{n}\times\R_{\geq0}$
(cf.\cite[p. 83, Theorem 5]{Filippov1988}).}

In this paper, the solutions to (\ref{eq:Inclusion}) are analyzed
using Lyapunov-like comparison functions with negative semidefinite
derivatives. To this end, generalized time derivatives and non-strict
Lyapunov functions are defined as follows.
\begin{defn}
\label{def:Generalized time derivative}\textcolor{red}{Let $F:\mathcal{\R}^{n}\times\R_{\geq t_{0}}\rightrightarrows\mathcal{\R}^{n}$
be nonempty and compact valued.} The generalized time derivative of
a locally Lipschitz-continuous function $V:\mathcal{\R}^{n}\times\R_{\geq t_{0}}\to\R$
with respect to $F$ is the function $\dot{\bar{V}}_{F}:\mathcal{\R}^{n}\times\R_{\geq t_{0}}\to\R$
defined as (cf.\cite{Teel2000c})
\begin{equation}
\dot{\bar{V}}_{F}\left(x,t\right)\coloneqq\max_{p\in\partial V\left(x,t\right)}\max_{q\in F\left(x,t\right)}p^{T}\left[q;1\right],\label{eq:generalized time derivative}
\end{equation}
where $\partial V$ denotes the Clarke gradient of $V$ \cite[p. 39]{Clarke1990}.\defnEnd
\end{defn}
\textcolor{blue}{See Section \ref{sec:Comments-on-the} for a detailed
comparison of Definition \ref{def:Generalized time derivative} with
more popular set-valued notions of generalized time derivatives (i.e.,
\cite[eq. 13]{Shevitz1994} and \cite[p. 364]{Bacciotti.Ceragioli1999}).}
\begin{defn}
\label{def:Semidefinite Lyapunov Function}Let $\mathcal{D}\subseteq\R^{n}$
be an open and connected set and let $\Omega\coloneqq\mathcal{D}\times\mathcal{I}$
for some interval $\mathcal{I}$. \textcolor{red}{Let $F:\mathcal{\R}^{n}\times\R_{\geq t_{0}}\rightrightarrows\mathcal{\R}^{n}$
be nonempty and compact valued over $\Omega$. Let $V:\Omega\to\R$
be a locally Lipschitz-continuous positive definite function. Let
$\overline{W}\:,\underline{W}:\mathcal{D}\to\R$ be continuous positive
definite functions and let $W:\mathcal{D}\to\R$ be a continuous positive
semidefinite function. If}
\begin{equation}
\underline{W}\left(x\right)\leq V\left(x,t\right)\leq\overline{W}\left(x\right),\quad\forall\left(x,t\right)\in\Omega,\label{eq:VBounds}
\end{equation}
and
\begin{align}
\dot{\bar{V}}_{F}\left(x,t\right) & \leq-W\left(x\right),\label{eq:VDecreases}
\end{align}
$\forall x\in\mathcal{D}$ and for almost all $t\in\R_{\geq t_{0}}$,
then $V$ is called a non-strict Lyapunov function for $F$ \textcolor{red}{over
$\Omega$ with the bounds $\underline{W}$, $\overline{W}$, and $W$.}\defnEnd
\end{defn}
The following theorem establishes the fact that the existence of a
non-strict Lyapunov function implies that $t\mapsto W\left(x\left(t\right)\right)$
asymptotically decays to zero. 
\begin{thm}
\label{thm:GLYT}Let $\mathcal{D}\subseteq\R^{n}$ be an open and
connected set, $r>0$ be selected such that $\overline{\b}\left(0,r\right)\subset\mathcal{D}$
and $\Omega\coloneqq\mathcal{D}\times\R_{\geq t_{0}}$. \textcolor{red}{Let
$F:\R^{n}\times\R_{\geq t_{0}}\rightrightarrows\R^{n}$ be a map that
satisfies Assumption \ref{ass:existence} and is locally bounded,
uniformly in $t$, over $\Omega$.}\footnote{A set valued map $F:\R^{n}\times\R_{\geq0}\rightrightarrows\R^{n}$
is locally bounded, uniformly in $t$, over $\Omega$, if for every
compact $K\subset\mathcal{D}$, there exists $M>0$ such that $\forall\left(x,t,y\right)$
such that $\left(x,t\right)\in K\times\R_{\geq t_{0}}$, and $y\in F\left(x,t\right)$,
$\left\Vert y\right\Vert _{2}\leq M$. } If $V:\Omega\to\R$ is a non-strict Lyapunov function for $F$ \textcolor{red}{over
$\Omega$ with the bounds $\underline{W}:\mathcal{D}\to\R$, $\overline{W}:\mathcal{D}\to\R$,
and $W:\mathcal{D}\to\R$}, then all solutions of (\ref{eq:Inclusion})
such that $x\left(t_{0}\right)\in\left\{ x\in\overline{\b}\left(0,r\right)\mid\overline{W}\left(x\right)\leq c\right\} $,
for some $c\in\left(0,\min_{\left\Vert x\right\Vert _{2}=r}\underline{W}\left(x\right)\right)$,
are complete, bounded, and satisfies $\lim_{t\to\infty}W\left(x\left(t\right)\right)=0$.
\textcolor{blue}{In addition, if $\mathcal{D}=\R^{n}$ and if the
sets $\left\{ x\in\R^{n}\mid\underline{W}\left(x\right)\leq c\right\} $
are compact, $\forall c\in\R_{>0}$, then the result is global. Furthermore,
if the non-strict Lyapunov function is regular \cite[Definition 2.3.4]{Clarke1990},
then (\ref{eq:VDecreases}) can be relaxed to $\dot{\underline{V}}_{F}\left(x,t\right)\leq-W\left(x\right)$,
where 
\begin{equation}
\dot{\underline{V}}_{F}\left(x,t\right)\coloneqq\min_{p\in\partial V\left(x,t\right)}\max_{q\in F\left(x,t\right)}p^{T}\left[q;1\right].\label{eq:VUnderlineDot}
\end{equation}
}
\end{thm}
\begin{IEEEproof}
See the appendix.
\end{IEEEproof}
\textcolor{blue}{The following section utilizes the results of Sections
}\ref{sec:Switching-and-regularization}\textcolor{blue}{{} and }\ref{sec:Semidefinite-Lyapunov-functions}\textcolor{blue}{{}
to develop the main results of this paper.}

\section{\label{sec:Main-Result}Invariance-like results for switched systems}

The following proposition states that a common non-strict Lyapunov
function for a family of differential inclusions is also a non-strict
Lyapunov function for the closure of their convex combination.\footnote{The observation that a common (strong) continuously differentiable
Lyapunov function for a family of finitely many differential inclusions
is also a Lyapunov function for the closure of their convex combination
is stated in \cite[Proposition 1]{Poveda.Teel2017}. In this paper,
it is proved and extended to families of countably infinite differential
inclusions and semidefinite locally Lipschitz-continuous Lyapunov
functions.}
\begin{prop}
\label{prop:Final}\textcolor{blue}{Let $\Omega\subseteq\R^{n}\times\R_{\geq t_{0}}$.
Let $\left\{ F_{\sigma}:\mathcal{\R}^{n}\times\R_{\geq t_{0}}\rightrightarrows\mathcal{\R}^{n}\mid\sigma\in\mathcal{N}^{o}\right\} $
be a family of set-valued maps with compact and nonempty values that
is locally bounded, uniformly in $\sigma$, over $\Omega\times\mathcal{N}^{o}$.}\footnote{\textcolor{blue}{\label{fn:regulardef}A collection of set valued
maps $\left\{ F_{\sigma}:\mathcal{\R}^{n}\times\R_{\geq t_{0}}\rightrightarrows\mathcal{\R}^{n}\mid\sigma\in\mathcal{N}^{o}\right\} $
is locally bounded, uniformly in $\sigma$, over $\Omega\times\mathcal{N}^{o}$,
if for every compact $K\subset\Omega$, there exists $M>0$ such that
$\forall\left(x,t,\sigma,y\right)$ such that $\left(x,t,\sigma\right)\in K\times\mathcal{N}^{o}$
and $y\in F_{\sigma}\left(x,t\right)$, $\left\Vert y\right\Vert _{2}\leq M$.}} If $V:\Omega\to\R$ is a common non-strict Lyapunov function for
the family $\left\{ F_{\sigma}\right\} $ over $\Omega$ \textcolor{red}{with
the bounds $\underline{W}:\mathcal{D}\to\R$, $\overline{W}:\mathcal{D}\to\R$,
and $W:\mathcal{D}\to\R$} (i.e., $V$ is a non-strict Lyapunov function
for $F_{\sigma}$ for each $\sigma\in\mathcal{N}^{o}$ and the bounds
$W$, $\overline{W}$, and $\underline{W}$ \ref{def:Semidefinite Lyapunov Function}
are independent of $\sigma$), then $V$ is also a non-strict Lyapunov
function for $\overline{\co}\bigcup_{\sigma\in\mathcal{N}^{o}}F_{\sigma}\left(x,t\right)$
over $\Omega$ \textcolor{red}{with the bounds $\underline{W}$, $\overline{W}$,
and $W$}.
\end{prop}
\begin{IEEEproof}
\textcolor{blue}{Since the maps $\left\{ F_{\sigma}\right\} $ are
locally bounded, uniformly in $\sigma$, over $\Omega\times\mathcal{N}^{o}$,
$\overline{\co}\bigcup_{\sigma\in\mathcal{N}^{o}}F_{\sigma}\left(x,t\right)$
is nonempty and compact for all $\left(x,t\right)\in\Omega$. Since
$V$ is a common non-strict Lyapunov function, $\max_{p\in V\left(x,t\right)}\max_{q\in F_{\sigma}\left(x,t\right)}p^{T}\left[q;1\right]\leq-W\left(x\right)$,
$\forall\text{\ensuremath{\sigma\in\mathcal{N}}}^{o}$. Let $q^{*}\in F\left(x,t\right)\coloneqq\overline{\co}\bigcup_{\sigma\in\mathcal{N}^{o}}F_{\sigma}\left(x,t\right)$.
There exists a sequence $\left\{ q_{j}\right\} _{j\in\n}$ such that
$\lim_{j\to\infty}q_{j}=q^{*}$ and $q_{j}\in\co\bigcup_{\sigma\in\mathcal{N}^{o}}F_{\sigma}\left(x,t\right)$.
By Carath\'{e}odory's theorem \cite[p. 103]{Danzer.Gruenbaum.ea1963},
$q_{j}=\sum_{i=1}^{m}a_{i}^{j}z_{i}^{j}$, where $\sum_{i=1}^{m}a_{i}^{j}=1$,
$a_{i}^{j}\geq0$, and $z_{i}^{j}\in F_{\sigma_{i}^{j}}\left(x,t\right),$
$\forall i\in\left\{ 1,\cdots,m\right\} $.}

\textcolor{blue}{For any fixed $p\in\partial V\left(x,t\right)$,
$p^{T}\left[z_{i}^{j};1\right]\leq\max_{q\in F_{\sigma_{i}^{j}}\left(x,t\right)}p^{T}\left[q;1\right],$
$\forall i\in\left\{ 1,\cdots,m\right\} $ and $\forall j\in\n$.
Hence,
\[
\max_{p\in\partial V\left(x,t\right)}p^{T}\left[z_{i}^{j};1\right]\!\leq\!\max_{p\in\partial V\left(x,t\right)}\max_{q\in F_{\sigma_{i}^{j}}\left(x,t\right)}p^{T}\left[q;1\right]\!\leq\!-W\left(x\right).
\]
$\forall i\in\left\{ 1,\cdots,m\right\} $ and $\forall j\in\n$.
Since $\sum_{i=1}^{m}a_{i}^{j}=1$, $\max_{p\in\partial V\left(x,t\right)}p^{T}\left[q_{j};1\right]\leq-W\left(x\right),$
$\forall j\in\n$. Now, since $\left(p,q\right)\mapsto p^{T}\left[q;1\right]$
is continuous, and $\partial V\left(x,t\right)$ and $\overline{\co}\bigcup_{\sigma\in\mathcal{N}^{o}}F_{\sigma}\left(x,t\right)$
are compact, the function $q\mapsto\max\left\{ p^{T}\left[q;1\right]\mid p\in\partial V\left(x,t\right)\right\} $
is continuous on $\overline{\co}\bigcup_{\sigma\in\mathcal{N}^{o}}F_{\sigma}\left(x,t\right)$.
Hence, $\max_{p\in\partial V\left(x,t\right)}p^{T}\left[q;1\right]\leq-W\left(x\right),$
$\forall q\in\overline{\co}\bigcup_{\sigma\in\mathcal{N}^{o}}F_{\sigma}\left(x,t\right)$.}
\end{IEEEproof}
\textcolor{blue}{The following corollary demonstrates that the bound
}(\ref{eq:VDecreases})\textcolor{blue}{{} in Proposition \ref{prop:Final}
can be relaxed to utilize $\dot{\underline{V}}_{F}$ instead of $\dot{\bar{V}}_{F}$
at the expense of a stricter continuity assumption on the set-valued
maps $\left\{ F_{\sigma}\right\} $.}
\begin{cor}
\textcolor{blue}{\label{cor:minmaxCts}Let the family of set-valued
maps $\left\{ F_{\sigma}:\mathcal{\R}^{n}\times\R_{\geq t_{0}}\rightrightarrows\mathcal{\R}^{n}\mid\sigma\in\mathcal{N}^{o}\right\} $
satisfy the hypothesis of Proposition \ref{prop:Final}. If $V:\Omega\to\R$
is a common non-strict regular Lyapunov function for the family $\left\{ F_{\sigma}\right\} $,
over $\Omega$, }\textcolor{red}{with the bounds $\underline{W}:\mathcal{D}\to\R$,
$\overline{W}:\mathcal{D}\to\R$, and $W:\mathcal{D}\to\R$,}\textcolor{blue}{{}
and with (\ref{eq:VDecreases}) in Definition \ref{def:Semidefinite Lyapunov Function}
relaxed to $\dot{\underline{V}}_{F_{\sigma}}\left(x,t\right)\leq-W\left(x\right)$,
$\forall\left(x,\sigma\right)\in\mathcal{\R}^{n}\times\mathcal{N}^{o}$
and for almost all $t\in\R_{\geq t_{0}}$, then $\dot{\underline{V}}_{\overline{\co}\bigcup_{\sigma\in\mathcal{N}^{o}}F_{\sigma}}\left(x,t\right)\leq-W\left(x\right)$,
$\forall\left(x,\sigma\right)\in\mathcal{\R}^{n}\times\mathcal{N}^{o}$
and for almost all $t\in\R_{\geq t_{0}}$, provided the set-valued
maps $\left\{ F_{\sigma}\right\} $ are continuous (in the sense of
\cite[Definition 1.4.3]{Aubin2008})}\textcolor{red}{{} and convex valued.}\textcolor{blue}{}\footnote{\textcolor{blue}{Example \ref{exa:Counterexample} demonstrates that
there are collections of upper semicontinuous set-valued maps for
which Corollary }\ref{cor:minmaxCts}\textcolor{blue}{{} fails to hold,
i.e., the continuity hypothesis in Corollary }\ref{cor:minmaxCts}\textcolor{blue}{{}
is not vacuous. }}
\end{cor}
\begin{IEEEproof}
\textcolor{blue}{See the appendix.}
\end{IEEEproof}
\textcolor{blue}{The main result of the paper can now be summarized
in the following theorem. }
\begin{thm}
\label{thm:FinalResult}Let $r>0$ be selected such that $\overline{\b}\left(0,r\right)\subset\mathcal{D}$
and let $\Omega\coloneqq\mathcal{D}\times\R_{\geq t_{0}}$. If \textcolor{blue}{Assumption
\ref{assu:locally-finite-switches} holds and} the (Filippov) Krasovskii
regularizations of the subsystems in (\ref{eq:Subsystems}) admit
a common non-strict Lyapunov \textcolor{red}{function $V:\Omega\to\R$,
over $\Omega$, with the bounds $\underline{W}:\mathcal{D}\to\R$,
$\overline{W}:\mathcal{D}\to\R$, and $W:\mathcal{D}\to\R$, }then
every solution of the (Filippov) Krasovskii regularization of the
switched system in (\ref{eq:Switched System}) such that $x\left(t_{0}\right)\in\left\{ x\in\b\left(0,r\right)\mid\overline{W}\left(x\right)\leq c\right\} $,
\textcolor{blue}{for some $c\in\left(0,\min_{\left\Vert x\right\Vert _{2}=r}\underline{W}\left(x\right)\right)$,}
is complete, bounded, and satisfies $\lim_{t\to\infty}W\left(x\left(t\right)\right)=0$.
\textcolor{blue}{In addition, if $\mathcal{D}=\R^{n}$ and if the
sets $\left\{ x\in\R^{n}\mid\underline{W}\left(x\right)\leq c\right\} $
are compact, $\forall c\in\R_{>0}$, then the result is global.}
\end{thm}
\begin{IEEEproof}
\textcolor{blue}{Since the collection $\left\{ f_{\sigma}\mid\sigma\in\mathcal{N}^{o}\right\} $
is locally bounded, uniformly in $t$ and $\sigma$, over $\Omega\times\mathcal{N}^{o}$,
the collections $\left\{ \mathbb{F}_{\sigma}\mid\sigma\in\mathcal{N}^{o}\right\} $
and $\left\{ \mathbb{K}_{\sigma}\mid\sigma\in\mathcal{N}^{o}\right\} $
are also locally bounded, uniformly in $t$ and $\sigma$, over $\Omega\times\mathcal{N}^{o}$.
Hence, by Proposition \ref{prop:Final}, $V$ is also a non-strict
Lyapunov function for the set-valued maps $\left(x,t\right)\mapsto\overline{\co}\bigcup_{\sigma\in\mathcal{N}^{o}}\mathbb{F}_{\sigma}\left(x,t\right)$
and $\left(x,t\right)\mapsto\overline{\co}\bigcup_{\sigma\in\mathcal{N}^{o}}\mathbb{K}_{\sigma}\left(x,t\right)$,
over $\Omega$, }\textcolor{red}{with the bounds $\underline{W}$,
$\overline{W}$, and $W$}\textcolor{blue}{. From Proposition \ref{prop:The-set-valued-maps},
$\mathbb{F}\left(x,t\right)\subseteq\overline{\co}\bigcup_{\sigma\in\mathcal{N}^{o}}\mathbb{F}_{\sigma}\left(x,t\right)$
and $\mathbb{K}\left(x,t\right)\subseteq\overline{\co}\bigcup_{\sigma\in\mathcal{N}^{o}}\mathbb{K}_{\sigma}\left(x,t\right)$.
Hence, $V$ is also a non-strict Lyapunov function for $\mathbb{F}$
and $\mathbb{K}$, over $\Omega$, }\textcolor{red}{with the bounds
$\underline{W}$, $\overline{W}$, and $W$}\textcolor{blue}{. Since
$f$ is locally bounded, uniformly in $t$ over $\Omega$, $\mathbb{F}$
and $\mathbb{K}$ are also locally bounded, uniformly in $t$ over
$\Omega$. The conclusion then follows by Theorem \ref{thm:GLYT}.}
\end{IEEEproof}
\begin{rem}
\label{rem:The-geometric-condition}The geometric condition in (\ref{eq:VDecreases})
can be relaxed to the following trajectory-based condition. For all
generalized solutions $x_{\sigma}:\mathcal{I}\to\R^{n}$ to (\ref{eq:Subsystems}),
let the subsystems in (\ref{eq:Subsystems}) satisfy 
\begin{equation}
\dot{\bar{V}}_{F_{\sigma}}\left(x_{\sigma}\left(t\right),t\right)\leq-W\left(x_{\sigma}\left(t\right)\right),\label{eq:Relaxed V bound}
\end{equation}
$\forall\sigma\in\mathcal{N}^{o}$ and for almost all $t\in\mathcal{I}$.
In addition, for a specific generalized solution $x^{*}:\mathcal{I}\to\R^{n}$
to (\ref{eq:Switched System}), if the set $\left\{ t\subseteq\mathcal{I}\mid\rho\left(x^{*}\left(\cdot\right),\cdot\right)\:\textnormal{is discontinuous at }t\right\} $
is countable for every $\mathcal{I}\subseteq\R_{\ge t_{0}}$, then
weak versions of Theorem \ref{thm:GLYT} and Proposition \ref{prop:Final}
that establish the convergence of $W\left(x^{*}\left(t\right)\right)$
to the origin as $t\to\infty$ can be proven using techniques similar
to \cite[Corollary 1]{Fischer.Kamalapurkar.ea2013}.
\end{rem}
\begin{rem}
\textcolor{blue}{If the subsystems are autonomous, and if they admit
a common non-strict Lyapunov function that is regular, then by applying
the invariance principle (e.g., \cite[Theorem 3]{Bacciotti.Ceragioli1999})
to the differential inclusions $\dot{x}\in\overline{\co}\bigcup_{\sigma\in\mathcal{N}^{o}}\mathbb{F}_{\sigma}\left(x\right)$
and $\dot{x}\in\overline{\co}\bigcup_{\sigma\in\mathcal{N}^{o}}\mathbb{K}_{\sigma}\left(x\right)$,
it can be shown that all solutions of (\ref{eq:Switched System})
that start in the set $L_{l}$ converge to the largest weakly forward
invariant set contained within $L_{l}\cap\left\{ x\in\mathcal{D}\mid W\left(x\right)=0\right\} $,
where $L_{l}$ is a bounded connected component of the level set $\left\{ x\in\mathcal{D}\mid V\left(x\right)\leq l\right\} $.
Hence, Propositions \ref{prop:The-set-valued-maps} and \ref{prop:Final}
also generalize results such as \cite{Bacciotti.Mazzi2005} to switched
nonsmooth systems. A similar result can also be obtained for the case
where the subsystems are periodic.}
\end{rem}

\section{\label{sec:Comments-on-the}Comments on the generalized time derivative}

\textcolor{blue}{If $V$ is regular then the generalized time derivative
obtained using Definition \ref{def:Generalized time derivative} is
generally more conservative than (i.e., greater than or equal to)
the maximal element of the more popular set-valued generalized derivatives
defined in \cite{Shevitz1994} and \cite{Bacciotti.Ceragioli1999}.
The motivation behind the use of the seemingly restrictive definition
is twofold: (a) through a reduction of the admissible directions in
$F$ using locally Lipschitz-continuous regular functions, a generalized
time derivative that is less conservative than the set-valued derivatives
in \cite{Shevitz1994} and \cite{Bacciotti.Ceragioli1999} can be
obtained (see Lemma \ref{lem:Reduction} and Corollary \ref{cor:ReducedStability})
and (b) the invariance-like results in Section \ref{sec:Main-Result}
do not hold if the time derivative of the cLf is interpreted in the
set-valued sense (see Example \ref{exa:Counterexample}).} 
\begin{lem}
\label{lem:Reduction}Let $\mathcal{D}\subseteq\R^{n}$ be open and
connected and let $\Omega\coloneqq\mathcal{D}\times\R_{\geq t_{0}}$.
Let $V:\Omega\to\R$ be a locally Lipschitz continuous function and
let $\mathcal{V}\coloneqq\left\{ V_{i}:\Omega\to\R\right\} _{i\in\mathcal{N}\subset\n}$
be a countable collection of locally Lipschitz-continuous regular
functions. \textcolor{red}{Let $F:\mathcal{\R}^{n}\times\R_{\geq t_{0}}\rightrightarrows\mathcal{\R}^{n}$
be a map that satisfies Assumption \ref{ass:existence} and} let $G,G_{i},\tilde{F}:\mathcal{\R}^{n}\times\R_{\geq t_{0}}\rightrightarrows\mathcal{\R}^{n}$
be defined as
\begin{align*}
G_{i}\left(x,t\right) & \coloneqq\!\left\{ \!q\in F\left(x,t\right)\!\mid\!\exists a_{f}\!\mid\!p^{T}\!\left[q;1\right]=a_{f},\forall p\in\partial V_{i}\left(x,t\right)\!\right\} \!,\\
G\left(x,t\right) & \coloneqq\!\left\{ \!q\in F\left(x,t\right)\!\mid\!\exists a_{f}\!\mid\!p^{T}\!\left[q;1\right]=a_{f},\forall p\in\partial V\left(x,t\right)\!\right\} \!,\\
\tilde{F}\left(x,t\right) & \coloneqq F\left(x,t\right)\cap\left(\cap_{i=1}^{\infty}G_{i}\left(x,t\right)\right),
\end{align*}
$\forall\left(x,t\right)\in\Omega$. If 
\begin{equation}
{\color{blue}\dot{\underline{V}}_{\tilde{F}}\left(x,t\right)}\leq-W\left(x\right),\quad\forall\left(x,t\right)\in\Omega,\label{eq:ReducedF}
\end{equation}
where $\dot{\underline{V}}_{\tilde{F}}$ is the $\mathcal{V}-$generalized
time derivative of $V$ with respect to $F:\mathcal{\R}^{n}\times\R_{\geq t_{0}}\rightrightarrows\mathcal{\R}^{n}$,
defined as 
\[
\dot{\underline{V}}_{\tilde{F}}\left(x,t\right)\coloneqq\min_{p\in\partial V\left(x,t\right)}\max_{q\in\tilde{F}\left(x,t\right)}p^{T}\left[q;1\right],
\]
$\forall\left(x,t\right)\in\Omega,$ and $\dot{\underline{V}}_{\tilde{F}}\left(x,t\right)$
is understood to be $-\infty$ when $\tilde{F}\left(x,t\right)$ is
empty, then each solution of (\ref{eq:Inclusion}), such that $x\left(t_{0}\right)\in\mathcal{D}$,
satisfies $\dot{V}\left(x\left(t\right),t\right)\leq-W\left(x\left(t\right)\right)$,
for almost all $t\in\left[t_{0},T\right)$, where $T\coloneqq\min\left(\sup\mathcal{I},\inf\left\{ t\in\mathcal{I}\mid x\left(t\right)\notin\mathcal{D}\right\} \right)$.
\end{lem}
\begin{IEEEproof}
See the appendix.
\end{IEEEproof}
Lemma \ref{lem:Reduction} implies that to establish Lyapunov stability
and asymptotic behavior of all solutions of (\ref{eq:Inclusion}),
examination of the set $\tilde{F}$, reduced from $F$ using the functions
in $\mathcal{V}$, is sufficient. In \cite[p. 364]{Bacciotti.Ceragioli1999},
the maximization is performed over the set $G$ instead of $\tilde{F}$,
i.e.,\footnote{The minimization here serves to maintain consistency of notation,
but is in fact, redundant.}
\[
\max\dot{\bar{V}}^{\left(F\right)}\left(x,t\right)=\min_{p\in\partial V\left(x,t\right)}\max_{q\in G\left(x,t\right)}p^{T}\left[q;1\right].
\]
\textcolor{blue}{Note that if $V\in\mathcal{V}$ then $\dot{\underline{V}}_{\tilde{F}}=\dot{\bar{V}}_{\tilde{F}}$
and $\tilde{F}\subseteq G$, and hence, $\dot{\underline{V}}_{\tilde{F}}\left(x,t\right)=\dot{\bar{V}}_{\tilde{F}}\left(x,t\right)\leq\max\dot{\bar{V}}^{\left(F\right)}\left(x,t\right),$
$\forall\left(x,t\right)\in\Omega$}. Thus, depending on the functions
$\mathcal{V}$ selected to reduce the inclusions, \textcolor{blue}{both
$\dot{\underline{V}}_{\tilde{F}}$ and $\dot{\bar{V}}_{\tilde{F}}$
can provide} a notion of the generalized time derivative of $V$ that
is less conservative than the set-valued derivative in \cite{Bacciotti.Ceragioli1999}
(and hence, the set-valued derivative in \textcolor{blue}{\cite{Shevitz1994}}).
Naturally, if $\mathcal{V}=\left\{ V\right\} $ then the three are
equivalent. The following definition is inspired by Lemma \ref{lem:Reduction}
and the corollary that follows is a straightforward consequence of
Theorem \ref{thm:GLYT} and Lemma \ref{lem:Reduction}.
\begin{defn}
\label{def:Reduced Semidefinite Lyapunov Function}Let $\mathcal{D}\subseteq\R^{n}$
be open and connected and let $\Omega\coloneqq\mathcal{D}\times\R_{\geq t_{0}}$.
Let $V:\Omega\to\R$ be a locally Lipschitz-continuous regular function
and \textcolor{red}{let $\overline{W},\underline{W}:\Omega\to\R$
be continuous positive definite functions that satisfy (\ref{eq:VBounds}).}
Let $\mathcal{V}$ be a countable collection of locally Lipschitz-continuous
regular functions. If there exists a continuous positive semidefinite
function $W:\mathcal{D}\to\R$ such that $\dot{\underline{V}}_{\tilde{F}}\left(x,t\right)\leq-W\left(x\right),$
$\forall x\in\mathcal{D}$ and for almost all $t\in\R_{\geq t_{0}}$,
then $V$ is called a $\mathcal{V}-$non-strict Lyapunov function
for $F:\mathcal{\R}^{n}\times\R_{\geq t_{0}}\rightrightarrows\mathcal{\R}^{n}$
\textcolor{red}{over $\Omega$ with the bounds $\overline{W}$, $\underline{W}$,
and $W$.}\defnEnd
\end{defn}
\begin{cor}
\label{cor:ReducedStability}Let $\mathcal{D}\subseteq\R^{n}$ be
an open and connected set containing the origin and let $\Omega\coloneqq\mathcal{D}\times\R_{\geq t_{0}}$.
Assume that the differential inclusion in (\ref{eq:Inclusion}) admits
a $\mathcal{V}-$non-strict Lyapunov function \textcolor{red}{over
$\Omega$ with the bounds $\overline{W}:\mathcal{D}\to\R$, $\underline{W}\mathcal{:D}\to\R$,
and $W:\mathcal{D}\to\R$. If $F:\R^{n}\times\R_{\geq t_{0}}\rightrightarrows\R^{n}$
satisfies Assumption \ref{ass:existence} and is }locally bounded,
uniformly in $t$, over $\Omega$, then every solution of (\ref{eq:Inclusion})
such that \textcolor{blue}{$x\left(t_{0}\right)\in\left\{ x\in\b\left(0,r\right)\mid\overline{W}\left(x\right)\leq c\right\} $,
for some $c\in\left(0,\min_{\left\Vert x\right\Vert _{2}=r}\underline{W}\left(x\right)\right)$,}
is complete, bounded, and satisfies $\lim_{t\to\infty}W\left(x\left(t\right)\right)=0$.\defnEnd
\end{cor}
At this juncture, it would be natural to ask whether the result in
Theorem \ref{thm:FinalResult} can be established using the\textcolor{blue}{{}
set-valued derivatives in \cite{Shevitz1994} and \cite{Bacciotti.Ceragioli1999}}
or a common $\mathcal{V}-$non-strict Lyapunov function. The following
example demonstrates that a common $\mathcal{V}-$non-strict Lyapunov
function is not sufficient to establish the results in Section \ref{sec:Main-Result}
and neither are the \textcolor{blue}{set-valued derivatives in \cite{Shevitz1994}
or \cite{Bacciotti.Ceragioli1999}}. Furthermore, the example also
demonstrates that the continuity assumption in Corollary \ref{cor:minmaxCts}
is not vacuous.
\begin{example}
\textcolor{blue}{\label{exa:Counterexample}Let $g_{1},g_{2},g_{3}:\R^{2}\to\R^{2}$
be defined as $g_{1}\left(x\right)\coloneqq\left[x_{1};0\right],$
$g_{2}\left(x\right)\coloneqq\left[0;x_{2}\right],$ and $g_{3}\left(x\right)\coloneqq\left[-x_{1};-x_{2}\right]$.
Let the subsystems be defined by $f_{1},f_{2}:\R^{2}\to\R^{2}$ as
\[
f_{1}\!\left(x\right)\!=\!\begin{cases}
\!g_{1}\left(x\right) & \!\!\left|x_{1}\right|\!<\left|x_{2}\right|\\
\!g_{3}\left(x\right) & \!\!\left|x_{1}\right|\geq\!\left|x_{2}\right|
\end{cases},\quad f_{2}\left(x\right)\!=\!\begin{cases}
\!g_{2}\left(x\right) & \!\!\left|x_{1}\right|\!<\left|x_{2}\right|\\
\!g_{3}\left(x\right) & \!\!\left|x_{1}\right|\geq\!\left|x_{2}\right|,
\end{cases}
\]
The subsystems have identical Krasovskii and Filippov regularizations,
given by
\begin{align*}
F_{1}\left(x\right) & =\begin{cases}
\overline{\co}\left\{ g_{1}\left(x\right),g_{3}\left(x\right)\right\}  & \left|x_{1}\right|=\left|x_{2}\right|\\
f_{1}\left(x\right) & \textnormal{otherwise},
\end{cases}\\
F_{2}\left(x\right) & =\begin{cases}
\overline{\co}\left\{ g_{2}\left(x\right),g_{3}\left(x\right)\right\}  & \left|x_{1}\right|=\left|x_{2}\right|\\
f_{2}\left(x\right) & \textnormal{otherwise}.
\end{cases}
\end{align*}
The function $V:\R^{2}\to\R$, defined as $V\left(x\right)\coloneqq\max\left(\left|x_{1}\right|,\left|x_{2}\right|\right)$,
is a locally Lipschitz-continuous regular function}\footnote{\textcolor{blue}{Pointwise maxima of locally Lipschitz-continuous
regular functions is locally Lipschitz-continuous and regular.}}\textcolor{blue}{{} that satisfies (\ref{eq:VBounds}) and
\[
\partial V\left(x\right)=\begin{cases}
v_{1}\left(x\right) & \left|x_{1}\right|<\left|x_{2}\right|\\
v_{2}\left(x\right) & \left|x_{1}\right|>\left|x_{2}\right|\\
\overline{\co}\left\{ v_{1}\left(x\right),v_{2}\left(x\right)\right\}  & \left|x_{1}\right|=\left|x_{2}\right|,
\end{cases}
\]
where $v_{1}\left(x\right)=\left[\sgn\left(x_{1}\right);0\right]$
and $v_{2}\left(x\right)=\left[0;\sgn\left(x_{2}\right)\right]$.
Hence, with $\mathcal{V}=\left\{ V\right\} $, $\tilde{F}_{i}\left(x\right)=\begin{cases}
\left\{ 0\right\}  & \left|x_{1}\right|=\left|x_{2}\right|\\
F_{i}\left(x\right) & \textnormal{otherwise}
\end{cases}$, for $i=1,2$.}

\textcolor{blue}{In this case, $\left(v_{1}\left(x\right)\right)^{T}f_{2}\left(x\right)=\left(v_{2}\left(x\right)\right)^{T}f_{1}\left(x\right)=0$,
$\left(v_{1}\left(x\right)\right)^{T}f_{3}\left(x\right)=-\left|x_{1}\right|$,
and $\left(v_{2}\left(x\right)\right)^{T}f_{3}\left(x\right)=-\left|x_{2}\right|$.
It follows that $\dot{\underline{V}}_{F_{i}}\left(x\right)\le0$ and
$\dot{\underline{V}}_{\tilde{F}_{i}}\left(x\right)=\dot{\bar{V}}_{\tilde{F}}\left(x\right)\le0$,
$\forall x\in\R^{2}$ and $i=1,2$. It is also easy to see that $\max\dot{\bar{V}}^{\left(F_{i}\right)}\left(x\right)\leq0$
and $\max\dot{\tilde{V}}^{\left(F_{i}\right)}\left(x\right)\leq0$,
$\forall x\in\R^{2}$ and $i=1,2$, where $\dot{\tilde{V}}^{\left(F_{i}\right)}$
is defined in \cite[eq. 13]{Shevitz1994}. Thus, $V$ is a common
non-strict cLf for the subsystems in the sense of $\dot{\underline{V}}_{F}$,
Definition \ref{def:Reduced Semidefinite Lyapunov Function}, \cite{Bacciotti.Ceragioli1999},
and \cite{Shevitz1994}.}

\textcolor{blue}{Let $F\coloneqq x\mapsto\overline{\co}F_{1}\left(x\right)\cup F_{2}\left(x\right)$.
For any $x\in\R^{2}$ such that $\left|x_{1}\right|=\left|x_{2}\right|$,
$q\coloneqq\frac{1}{2}\left[x_{1};x_{2}\right]\in\overline{\co}\left\{ g_{1}\left(x\right),g_{2}\left(x\right),g_{3}\left(x\right)\right\} =F\left(x\right)$.
Thus, whenever $\left|x_{1}\right|=\left|x_{2}\right|=V\left(x\right)>0$,
$\min_{p\in\partial V\left(x\right)}p^{T}q=0.5V\left(x\right)>0$,
i.e., Proposition \ref{prop:Final} does not hold. Furthermore, a
solution of $\dot{x}\in F\left(x\right)$, starting at $x=\left[1;1\right]$,
is $x\left(t\right)=\e^{0.5t}\left[1;1\right]$, i.e., Theorem \ref{thm:FinalResult}
does not hold.}
\end{example}
\textcolor{blue}{Thus, Proposition \ref{prop:Final} and Theorem \ref{thm:FinalResult}
may not hold if the generalized time derivative is understood in the
sense of Definition \ref{def:Reduced Semidefinite Lyapunov Function},
\cite{Bacciotti.Ceragioli1999} or \cite{Shevitz1994}. If $\dot{\underline{V}}_{F}$
is used as the generalized time derivative instead of $\dot{\bar{V}}_{F}$
then Proposition \ref{prop:Final} may not hold if the set-valued
maps $\left\{ F_{\sigma}\right\} $ are not continuous.\defnEnd}

\section{\label{sec:Example}Design Examples}

Many of the applications discussed in the opening paragraphs of Section
\ref{sec:Introduction} can be represented by the following example
problems. The first example demonstrates the utility of the developed
technique on a simple problem where only the regression dynamics are
discontinuous. In the second example, an adaptive controller for a
switched system that exhibits arbitrary switching between subsystems
with different parameters and disturbances is analyzed. 
\begin{example}
\label{exa:Simple}Consider the nonlinear dynamical system
\begin{equation}
\dot{x}=Y_{\rho\left(x,t\right)}\left(x\right)\theta+u+d\left(t\right),\label{eq:Ex1}
\end{equation}
where $x\in\R^{n}$ denotes the state, $u\in\R^{n}$ denotes the control
input, $d:\R_{\geq t_{0}}\to\R^{n}$ denotes an unknown disturbance,
$\rho:\R^{n}\times\R_{\geq t_{0}}\to\mathbb{N}$ denotes the switching
signal, $Y_{\sigma}:\R^{n}\to\R^{n\times L}$, for each $\sigma\in\mathbb{N}$,
is a known continuous function, and $\theta\in\R^{L}$ is the vector
of constant unknown parameters. The control objective is to regulate
the system state to the origin. The disturbance is assumed to be bounded,
with a known bound $\overline{d}$ such that $\left\Vert d\left(t\right)\right\Vert _{\infty}\leq\overline{d}$,
for almost all $t\in\R_{\ge t_{0}}$. Furthermore, $t\mapsto d\left(t\right)$
is assumed to be Lebesgue measurable.
\end{example}
The adaptive controller designed to satisfy the control objective
is $u=-kx-Y_{\rho\left(x,t\right)}\left(x\right)\hat{\theta}-\beta\sgn\left(x\right),$
where $\hat{\theta}:\R_{\geq t_{0}}\to\R^{L}$ denotes an estimate
of the vector of unknown parameters, $\theta$, $k,$ $\beta\in\R_{>0}$
are positive constant control gains, and $\sgn\left(\cdot\right)$
is the signum function. The estimate, $\hat{\theta}$, is obtained
from the update law $\dot{\hat{\theta}}=\left(Y_{\rho\left(x,t\right)}\left(x\right)\right)^{T}x.$
For each $\sigma\in\mathbb{N}$, the closed-loop error system can
then be expressed as 
\begin{align}
\dot{x} & =-kx+Y_{\sigma}\left(x\right)\tilde{\theta}+d\left(t\right)-\beta\sgn\left(x\right),\label{eq:xDotCL}\\
\dot{\tilde{\theta}} & =-\left(Y_{\sigma}\left(x\right)\right)^{T}x,\label{eq:ThetaTildeDotCL}
\end{align}
where $\tilde{\theta}\coloneqq\theta-\hat{\theta}$ denotes the parameter
estimation error. The closed-loop system in (\ref{eq:xDotCL}) and
(\ref{eq:ThetaTildeDotCL}) is discontinuous, and hence, does not
admit classical solutions. Thus, the analysis will focus on generalized
solutions to (\ref{eq:xDotCL}) and (\ref{eq:ThetaTildeDotCL}). Since
Filippov and Krasovskii regularizations of the closed-loop system
in (\ref{eq:xDotCL}) and (\ref{eq:ThetaTildeDotCL}) are identical,
the solutions to the corresponding differential inclusions are hereafter
simply referred to as generalized solutions.

To analyze the developed controller, consider the cLf $V:\R^{n+L}\to\R_{\geq t_{0}}$,
defined as
\begin{equation}
V\left(z\right)\coloneqq\frac{1}{2}x^{T}x+\frac{1}{2}\tilde{\theta}^{T}\tilde{\theta},\label{eq:V}
\end{equation}
where $z\coloneqq\left[x;\tilde{\theta}\right]$. Since the cLf is
continuously differentiable, the Clarke gradient reduces to the standard
gradient, i.e, $\partial V\left(z,t\right)=\left\{ z\right\} $. Using
the calculus of $\k\left[\cdot\right]$ from \cite{Paden1987}, a
bound on the regularization of the system in (\ref{eq:xDotCL}) and
(\ref{eq:ThetaTildeDotCL}) can be computed as $F_{\sigma}\left(z,t\right)\subseteq F_{\sigma}^{\prime}\left(z,t\right)$,
where
\[
F_{\sigma}^{\prime}\left(z,t\right)=\begin{bmatrix}\left\{ -kx+Y_{\sigma}\left(x\right)\tilde{\theta}+d\left(t\right)\right\} -\beta\k\left[\sgn\right]\left(x\right)\\
\left\{ -Y_{\sigma}^{T}\left(x\right)x\right\} 
\end{bmatrix}.
\]
Using the Definition \ref{def:Generalized time derivative} and the
fact that $x^{T}\k\left[\sgn\right]\left(x\right)=\left\{ \left\Vert x\right\Vert _{1}\right\} $,
a bound on the generalized time derivative of the cLf can be computed
as 
\begin{align*}
\dot{\bar{V}}_{\sigma}\left(z,t\right) & =\max_{q\in F_{\sigma}\left(z,t\right)}z^{T}q,\\
 & \leq\max_{q\in F_{\sigma}^{\prime}\left(z,t\right)}z^{T}q,\\
 & =-k\left\Vert x\right\Vert _{2}^{2}+x^{T}d\left(t\right)-\beta\left\Vert x\right\Vert _{1}.
\end{align*}
Provided $\beta>\overline{d}$, 
\begin{equation}
\dot{\bar{V}}_{\sigma}\left(z,t\right)\leq-W\left(z\right),\label{eq:VDotBound}
\end{equation}
$\forall\left(z,\sigma\right)\in\R^{n+L}\times\mathbb{N}$ and for
almost all $t\in\R_{\geq t_{0}}$, where $W\left(z\right)=k\left\Vert x\right\Vert _{2}^{2}$
is a positive semidefinite function. Using (\ref{eq:V}), (\ref{eq:VDotBound}),
and Theorem \ref{thm:FinalResult}, all the generalized solutions
of the switched nonsmooth system in (\ref{eq:xDotCL}) and (\ref{eq:ThetaTildeDotCL})
are complete, bounded, and satisfy $\left\Vert x\left(t\right)\right\Vert _{2}\to0$
as $t\to\infty$.\defnEnd
\begin{example}
Arbitrary switching between systems with different parameters and
disturbances can be achieved in the case where the number of subsystems
is finite. For example, consider the nonlinear dynamical system 
\begin{equation}
\dot{x}=Z_{\rho\left(x,t\right)}\left(x,t\right)\theta_{\rho\left(x,t\right)}+d_{\rho\left(x,t\right)}\left(x,t\right)+u,\label{eq:Ex2}
\end{equation}
where $\rho:\R^{n}\times\R_{\geq t_{0}}\to\mathcal{N}^{o}$ such that
$\mathcal{N}^{o}$ is a finite set, $Z_{\sigma}:\R^{n}\times\R_{\geq t_{0}}\to\R^{n\times L}$,
are known functions, $\theta_{\sigma}\in\R^{L}$ are vectors of constant
unknown parameters corresponding to each $\sigma\in\mathcal{N}^{o}$,
and $d_{\sigma}:\R^{n}\times\R_{\geq t_{0}}\to\R^{n}$ are unknown
disturbances such that for each $\sigma\in\mathcal{N}^{o}$, $\left\Vert d_{\sigma}\left(x,t\right)\right\Vert _{\infty}\leq\overline{d}_{\sigma},$
$\forall\left(x,t\right)\in\R^{n}\times\R_{\geq t_{0}}$ and some
$\overline{d}_{\sigma}>0$. Furthermore, for each $\sigma\in\R^{n}$,
$\left(x,t\right)\mapsto d_{\sigma}\left(x,t\right)$ and $\left(x,t\right)\mapsto Z_{\sigma}\left(x,t\right)$
are continuous in $x$, uniformly in $t$ and Lebesgue measurable
in $t$, $\forall x\in\R^{n}$. Let $\theta\coloneqq\left[\theta_{1};\theta_{2};\cdots;\theta_{\left|\mathcal{N}^{o}\right|}\right]\in\R^{L\left|\mathcal{N}^{o}\right|}$
and let $Y_{\sigma}\coloneqq\mathbf{1}_{\sigma}\otimes Z_{\sigma}$,
where $\mathbf{1}_{\sigma}\in\R^{1\times L}$ is a matrix defined
by 
\[
\left(\mathbf{1}_{\sigma}\right)_{1,j}=\begin{cases}
1, & j=\sigma.\\
0, & \text{otherwise}.
\end{cases}
\]
The adaptive controller designed to satisfy the control objective
is 
\[
u=-k_{\rho\left(x,t\right)}x-Y_{\rho\left(x,t\right)}\left(x,t\right)\hat{\theta}-\beta_{\rho\left(x,t\right)}\sgn\left(x\right),
\]
where $\beta_{\sigma}\in\R_{>0}$ and $k_{\sigma}\in\R_{>0}$ are
control gains corresponding to $\sigma\in\mathcal{N}^{o}$ and $\hat{\theta}:\R_{\geq t_{0}}\to\R^{L\left|\mathcal{N}^{o}\right|}$
is updated according to $\dot{\hat{\theta}}=\left(Y_{\rho\left(x,t\right)}\left(x,t\right)\right)^{T}x.$
A stability analysis similar to Example \ref{exa:Simple} can then
be utilized to conclude asymptotic convergence of the state $x$ to
the origin provided $\beta_{\sigma}>\overline{d}_{\sigma},$ $\forall\sigma\in\mathcal{N}^{o}$.\defnEnd
\end{example}

\section{\label{sec:Conclusion}Conclusion}

Motivated by applications in switched adaptive control, the generalized
LaSalle-Yoshizawa corollary in \cite{Fischer.Kamalapurkar.ea2013}
is extended to switched nonsmooth systems. The extension facilitates
the analysis of the asymptotic characteristics of a switched system
based on the asymptotic characteristics of the individual subsystems
where a non-strict common Lyapunov function can be constructed for
the subsystems. Application of the developed extension to a switched
adaptive system is demonstrated through simple examples. \textcolor{blue}{Motivated
by results such as \cite{Chen.Yangtoappaer}, further research could
potentially extend the developed method to utilize indefinite Lyapunov
functions.}

In Lemma \ref{lem:Reduction}, it is shown that arbitrary locally
Lipschitz-continuous regular functions can be used to reduce the differential
inclusion to a smaller set of admissible directions. This observation
indicates that there may be a smallest set of admissible directions
corresponding to each differential inclusion. Further research is
needed to establish the existence of such a set and to find a representation
of it that facilitates computation.

The developed method requires a strong convergence result for the
subsystems, i.e., the existence of a common cLf that satisfies (\ref{eq:VDecreases})
implies that all the generalized solutions to the individual subsystems
are bounded and asymptotically converge to the origin. Future research
will focus on the development of results for switched nonsmooth systems
where only weak convergence results (that is, only a subset of the
generalized solutions to the individual subsystems are bounded and
asymptotically converge to the origin) are available for the subsystems. 

\bibliographystyle{IEEEtran}
\bibliography{ncr,encr,master}

\appendix{}
\begin{IEEEproof}[Proof of Theorem \ref{thm:GLYT}]
Similar to the proof of \cite[Corollary 1]{Fischer.Kamalapurkar.ea2013},
\textcolor{blue}{it is established that the bounds on $\dot{\bar{V}}_{F}$
in (\ref{eq:generalized time derivative}) and }(\ref{eq:generalized time derivative})\textcolor{blue}{{}
imply that the cLf is nonincreasing along all the solutions of (\ref{eq:Inclusion}).
The nonincreasing property of the cLf is used to establish boundedness
of $x$, which is used to prove the existence and uniform continuity
of complete solutions. }Barb{\u{a}}lat\textcolor{blue}{'s lemma \cite[Lemma 8.2]{Khalil2002}
is then used to conclude the proof.}

\textcolor{blue}{To show that the cLf is nonincreasing,} let $x:\mathcal{I}\to\R^{n}$
be a maximal solution \textcolor{blue}{\cite[Definition 2.1]{Ryan1998}}
of (\ref{eq:Inclusion}) such that $x\left(t_{0}\right)\in\varOmega_{c}\coloneqq\left\{ x\in\overline{\b}\left(0,r\right)|\overline{W}\left(x\right)\leq c\right\} $.\textcolor{red}{}\footnote{\textcolor{red}{Similar to \cite[Proposition 1]{Ryan1990}, it can
be shown that any solution of (\ref{eq:Inclusion}) can be extended
to a maximal solution; hence, if a solution exists, then it can be
assumed to be maximal without loss of generality.}} Define $T>t_{0}$ be the first exit time of $x$ from $\mathcal{D}$,\textit{
}i.e., $T\coloneqq\min\left(\sup\mathcal{I},\inf\left\{ t\in\mathcal{I}\mid x\left(t\right)\notin\mathcal{D}\right\} \right)$,
where $\inf\emptyset$ is assumed to be $\infty$. If $V$ is locally
Lipschitz-continuous but not regular, then \cite[Proposition 4]{Ceragioli1999}
(see also, \cite[Theorem 2]{Moreau.Valadier1987}) can be used to
conclude that, for almost every $t\in\left[t_{0},T\right)$, \textcolor{blue}{the
time derivative $\dot{V}\left(x\left(t\right),t\right)$ exists,}
and $\exists p_{0}\in\partial V\left(x\left(t\right),t\right)$ such
that $\dot{V}\left(x\left(t\right),t\right)=p_{0}^{T}\left[\dot{x}\left(t\right);1\right]$.
Thus, (\ref{eq:generalized time derivative}) and (\ref{eq:VDecreases})
imply that $\dot{V}\left(x\left(t\right),t\right)\leq-W\left(x\left(t\right)\right)$
for almost every $t\in\left[t_{0},T\right)$. If $V$ is regular,
then the relaxation in Footnote \ref{fn:regulardef} and \cite[Equation 22]{Shevitz1994}
can be used to conclude that for almost every $t\in\left[t_{0},T\right)$,
\textcolor{blue}{the time derivative $\dot{V}\left(x\left(t\right),t\right)$
exists and} $\dot{V}\left(x\left(t\right),t\right)\leq-W\left(x\left(t\right)\right)$.
The conclusion that 
\begin{equation}
V\left(x\left(t_{0}\right),t_{0}\right)\geq V\left(x\left(t\right),t\right),\quad\forall t\in\left[t_{0},T\right)\label{eq:vdecreases}
\end{equation}
then follows from \cite[Lemma 2]{Fischer.Kamalapurkar.ea2013}.

\textcolor{blue}{Using (\ref{eq:vdecreases}), it can be shown that
(see, e.g., \cite[Theorem 4.8]{Khalil2002}) every solution of (\ref{eq:Inclusion})
that starts in $\varOmega_{c}$ stays in $\overline{\b}\left(0,r\right)$
on every interval of its existence. Therefore, all maximal solutions
of (\ref{eq:Inclusion}) such that $x\left(t_{0}\right)\in\varOmega_{c}$
are precompact \cite[Definition 2.3]{Ryan1998} and $T=\sup\mathcal{I}$.
}\textcolor{red}{In the following, arguments similar to \cite[Proposition 2]{Ryan1990}
are used to show that precompact solutions are complete.}

\textcolor{red}{For the sake of contradiction, assume that $T<\infty$.}
Since $F$ is locally bounded, uniformly in $t$, over $\Omega$,
and $x\left(t\right)\in\overline{\b}\left(0,r\right)$ on $\left[t_{0},T\right)$,
the map $t\mapsto F\left(x\left(t\right),t\right)$ is uniformly bounded
\textcolor{red}{on $\left[t_{0},T\right)$.} Hence, (\ref{eq:GenSol})
implies that \textcolor{red}{$\dot{x}\in\mathcal{L}_{\infty}\left(\left[t_{0},T\right),\R^{n}\right)$}.
Since $t\mapsto x\left(t\right)$ is locally absolutely continuous,
\textcolor{red}{$\forall t_{1},t_{2}\in\left[t_{0},T\right)$,} $\left\Vert x\left(t_{2}\right)-x\left(t_{1}\right)\right\Vert _{2}=\left\Vert \int_{t_{1}}^{t_{2}}\dot{x}\left(\tau\right)\d\tau\right\Vert _{2}$.
Since\textcolor{red}{{} $\dot{x}\in\mathcal{L}_{\infty}\left(\left[t_{0},T\right),\R^{n}\right)$,}
$\left\Vert \int_{t_{1}}^{t_{2}}\dot{x}\left(\tau\right)\d\tau\right\Vert _{2}\leq\int_{t_{1}}^{t_{2}}M\d\tau$,
where $M$ is a positive constant. Thus, $\left\Vert x\left(t_{2}\right)-x\left(t_{1}\right)\right\Vert _{2}\leq M\left|t_{2}-t_{1}\right|$,
and hence, $t\mapsto x\left(t\right)$ is uniformly continuous \textcolor{red}{on
$\left[t_{0},T\right)$. Therefore, $x$ can be extended into a continuous
function $x^{\prime}:\left[t_{0},T\right]\to\R^{n}$. Invoking \cite[p. 83, Theorem 5]{Filippov1988},
$x^{\prime}$ can be extended into a solution of (\ref{eq:Inclusion})
on the interval $\ropen{t_{0},T^{\prime}}$ for some $T^{\prime}>T$,
which contradicts the maximality of $x$. Hence, $T=\infty$, i.e.,
all precompact solutions of (\ref{eq:Inclusion}) are complete.}

Since $x\mapsto W\left(x\right)$ is continuous and $\overline{\b}\left(0,r\right)$
is compact, $x\mapsto W\left(x\right)$ is uniformly continuous on
$\overline{\b}\left(0,r\right)$. Since $t\mapsto x\left(t\right)$
is uniformly continuous on $\R_{\geq t_{0}}$, $t\mapsto W\left(x\left(t\right)\right)$
is uniformly continuous on $\R_{\geq t_{0}}$. Furthermore, $t\mapsto\intop_{t_{0}}^{t}W\left(x\left(\tau\right)\right)\d\tau$
is monotonically increasing and from (\ref{eq:VDecreases}) and the
fact that $V$ is positive definite,
\[
\intop_{t_{0}}^{t}\!\!W\!\left(x\left(\tau\right)\right)\d\tau\!\leq\!V\!\left(x\left(t_{0}\right),t_{0}\right)-V\!\left(x\left(t\right),t\right)\!\leq\!V\!\left(x\left(t_{0}\right),t_{0}\right).
\]
Hence, $\lim_{t\to\infty}\intop_{t_{0}}^{t}W\left(x\left(\tau\right)\right)\d\tau$
exists and is finite. By Barb{\u{a}}lat's Lemma \cite[Lemma 8.2]{Khalil2002},
$\lim_{t\to\infty}W\left(x\left(t\right)\right)=0$.
\end{IEEEproof}
\bigskip{}
\begin{IEEEproof}[Proof of Proposition \ref{prop:The-set-valued-maps} for Filippov
regularization]
\textcolor{blue}{Fix $\left(x,t\right)\in\R^{n}\times\R_{\geq t_{0}}$,
select $\delta^{*}>0$ such that $\left|\rho\left(\textnormal{B}\left(x,\delta^{*}\right),t\right)\right|<\infty$,
and let $\mathcal{N}\coloneqq\rho\left(\textnormal{B}\left(x,\delta^{*}\right),t\right)$.
Similar to the proof for Krasovskii regularization, the proof proceeds
in three steps. First, it is observed that 
\begin{multline}
\bigcap_{\delta>0}\bigcap_{\mu\left(N\right)=0}\overline{\co}\bigl\{ f_{\rho\left(y,t\right)}\left(y,t\right)\mid y\in\b\left(x,\delta\right)\setminus N\bigr\}\\
\subseteq\bigcap_{\delta>0}\bigcap_{\mu\left(N\right)=0}A_{N}^{\delta}\left(x,t\right),\label{eq:Filippov1}
\end{multline}
where $A_{N}^{\delta}\coloneqq\overline{\co}\bigcup_{\sigma\in\mathcal{N}}\bigl\{ f_{\sigma}\left(y,t\right)\mid y\in\b\left(x,\delta\right)\setminus N\bigr\}$.
Second, it is established that 
\begin{equation}
\bigcap_{\delta>0}\bigcap_{\mu\left(N\right)=0}A_{N}^{\delta}\left(x,t\right)\subseteq\bigcap_{\delta>0}\bigcap_{\mu\left(N\right)=0}B_{N}^{\delta}\left(x,t\right).\label{Filippov 2}
\end{equation}
 where $B_{N}^{\delta}\left(x,t\right)\coloneqq\co\bigcup_{\sigma\in\mathcal{N}}B_{N\delta\sigma}\left(x,t\right)$
and $B_{N\delta\sigma}\left(x,t\right)\coloneqq\overline{\co}\left\{ f_{\sigma}\left(y,t\right)\mid y\in\b\left(x,\delta\right)\setminus N\right\} .$
Finally, it is shown that $\forall x\in\R^{n}$ and almost all $t\in\R_{\geq t_{0}}$,
\begin{equation}
\bigcap_{\delta>0}\bigcap_{\mu\left(N\right)=0}B_{N}^{\delta}\left(x,t\right)\subseteq\co\bigcup_{\sigma\in\mathcal{N}}\bigcap_{\delta>0}\bigcap_{\mu\left(N\right)=0}B_{N\delta\sigma}\left(x,t\right).\label{eq:3F}
\end{equation}
The conclusion of the proposition then follows. Apart from the technical
detail required to handle the exclusion of measure-zero sets in the
Filippov inclusion, the methods utilized to prove  (\ref{Filippov 2})
and (\ref{eq:3F}) are similar to those used in the proof for Krasovskii
inclusions. Thus, in the following, only the techniques used to handle
the exclusion of measure-zero sets are illustrated.}

\textcolor{blue}{The containment in }(\ref{eq:Filippov1})\textcolor{blue}{{}
is self-evident. To prove (\ref{Filippov 2}),} define $\overline{\mathscr{N}}\left(\delta\right)\coloneqq\left\{ N\subset\b\left(x,\delta\right)\mid\mu\left(N\right)=0\right\} $,
and let $N^{*}\left(\delta\right)\subset2^{\b\left(x,\delta\right)}$
be a collection of sets of zero measure such that\textcolor{blue}{{}
$\sup\left\{ \left\Vert \theta\right\Vert \mid\theta\in A_{N}^{\delta}\right\} <\infty$,}
$\forall N\in N^{*}\left(\delta\right)$. Since the functions $f_{\sigma}\left(x,t\right)$
are locally essentially bounded, uniformly in $t$ and $\sigma$,
the collection $N^{*}\left(\delta\right)$ is nontrivial. Fix $N\in N^{*}\left(\delta\right)$
and $z\in A_{N}^{\delta}$. Using arguments similar to Part 1 of the
proof it can be shown that the point $z$ is a convex combination
of points from $B_{N\delta\sigma_{j}}\left(x,t\right)$. That is,
$z\in\co B_{N}^{\delta}\left(x,t\right)$, and hence, 
\begin{equation}
\bigcap_{N\in N^{*}\left(\delta\right)}A_{N}^{\delta}\left(x,t\right)\subseteq\bigcap_{N\in N^{*}\left(\delta\right)}B_{N}^{\delta}\left(x,t\right).\label{eq:FilippovIntermediate 1}
\end{equation}
\textcolor{blue}{To establish (\ref{Filippov 2}) the intersection
in (\ref{eq:FilippovIntermediate 1}) needs to include all of $\overline{\mathscr{N}}\left(\delta\right)$,
not just the subset $N^{*}\left(\delta\right)$.} Since $N^{*}\left(\delta\right)\subset\overline{\mathscr{N}}\left(\delta\right)$,
the inclusion $\bigcap_{N\in N^{*}\left(\delta\right)}A_{N}^{\delta}\left(x,t\right)\subseteq\bigcap_{N\in\overline{\mathscr{N}}\left(\delta\right)}A_{N}^{\delta}\left(x,t\right)$
follows. Let $M\in\overline{\mathscr{N}}\left(\delta\right)$. There
exist $N^{1}\in\overline{\mathscr{N}}\left(\delta\right)\setminus N^{*}\left(\delta\right)$
and $N^{0}\in N^{*}\left(\delta\right)$ such that $M=N^{1}\cup N^{0}$.
Since $N^{0}\subseteq M$, $A_{M}^{\delta}\left(x,t\right)\subset A_{N^{0}}^{\delta}\left(x,t\right)$.
Therefore, $\bigcap_{N\in\overline{\mathscr{N}}\left(\delta\right)}A_{N}^{\delta}\left(x,t\right)\subseteq\bigcap_{N\in N^{*}\left(\delta\right)}A_{N}^{\delta}\left(x,t\right)$,
which implies $\bigcap_{N\in N^{*}\left(\delta\right)}A_{N}^{\delta}\left(x,t\right)=\bigcap_{N\in\overline{\mathscr{N}}\left(\delta\right)}A_{N}^{\delta}\left(x,t\right)$.
A similar reasoning for $B_{N}^{\delta}\left(x,t\right)$ yields $\bigcap_{N\in N^{*}\left(\delta\right)}B_{N}^{\delta}\left(x,t\right)=\bigcap_{N\in\overline{\mathscr{N}}\left(\delta\right)}B_{N}^{\delta}\left(x,t\right)$,
\textcolor{blue}{$\forall\delta\in\left(0,\delta^{*}\right]$, }which
proves (\ref{Filippov 2}).

\textcolor{blue}{As an intermediate step towards proving (\ref{eq:3F}),
the containment 
\begin{equation}
\bigcap_{\mu\left(N\right)=0}\!\!\!\!B_{N}^{\delta}\left(x,t\right)\subseteq\co\!\!\bigcup_{\sigma\in\mathcal{N}}\bigcap_{\mu\left(N\right)=0}\!\!\!\!B_{N\delta\sigma}\left(x,t\right),\quad\forall\delta\in\left(0,\overline{\delta}\right],\label{eq:FilippovIntermediate 2}
\end{equation}
is established in the following.} Let $z\in\bigcap_{\mu\left(N\right)=0}B_{N}^{\delta}\left(x,t\right)$.
The objective now is to show that $z\in\co\left(\bigcap_{\mu\left(N\right)=0}B_{N\delta1}\left(x,t\right)\cup\bigcap_{\mu\left(N\right)=0}B_{N\delta2}\left(x,t\right)\cup\cdots\right).$
The inclusions in (\ref{eq:Filippov1}) and (\ref{Filippov 2}) are
valid $\forall\left(x,t\right)\in\R^{n}\times\R_{\geq t_{0}}$. For
the development hereafter, $\left(x,t\right)$ is restricted to a
set $\R^{n}\times E$ for some $E\subseteq\R_{\geq t_{0}}$ such that
$\forall\sigma\in\mathcal{N}$, the Filippov inclusions $\mathbb{F}_{\sigma}\left(x,t\right)$
can be expressed as $\bigcap_{\delta>0}\bigcap_{i\in\mathbb{N}}B_{N_{\sigma i}\delta\sigma}$
for some countable collection of measure zero sets $\left\{ N_{\sigma i}\left(t\right)\right\} _{i\in\mathbb{N}}$.
Under the additional assumption in (\ref{eq:FilippovCountableAssumption}),
the set $E$ can be selected to be equal to $\R_{\geq t_{0}}$. Define
$N^{*}\coloneqq\bigcup_{\sigma\in\mathcal{N}}\bigcup_{i\in\mathbb{N}}N_{\sigma i}$.
Since $N^{*}$ is a countable union of measure-zero sets, $\mu\left(N^{*}\right)=0$.
Since $z\in\bigcap_{\mu\left(N\right)=0}B_{N}^{\delta}\left(x,t\right)$,
by Carath\'{e}odory's Theorem \cite[p. 103]{Danzer.Gruenbaum.ea1963},
there exist $\left\{ z_{1},\cdots,z_{m}\right\} $ such that each
$z_{j}\in B_{N^{*}\delta\sigma_{j}}\left(x,t\right)$ for some $\sigma_{j}\in\mathcal{N}$,
and positive real numbers $\left\{ a_{1},\cdots,a_{m}\right\} $ with
$\sum_{j=1}^{m}a_{j}=1$, such that $z=\sum_{j=1}^{m}a_{j}z_{j}$.
Using (\ref{eq:FilippovCountableAssumption}) and De-Morgan's laws,
$B_{N^{*}\delta\sigma}\left(x,t\right)\subseteq\bigcap_{\mu\left(N\right)=0}B_{N\delta\sigma}\left(x,t\right)$,
$\forall\sigma\in\mathcal{N}$. \textcolor{blue}{Hence, for each $j\in\left\{ 1,\cdots,m\right\} $,
$z_{j}\in\bigcap_{\mu\left(N\right)=0}B_{N\delta\sigma_{j}}\left(x,t\right)$
for some $\sigma_{j}\in\mathcal{N}$, which implies (\ref{eq:FilippovIntermediate 2})},
\textcolor{blue}{$\forall\delta\in\left(0,\overline{\delta}\right]$.}
Using a nesting argument similar to the proof for Krasovskii inclusions,
the containment in (\ref{eq:3F}) follows $\forall\left(x,t\right)\in\R_{n}\times E$.

\textcolor{blue}{To complete the proof of (\ref{eq:3F}), it needs
to be established that }even without the additional assumption in
(\ref{eq:FilippovCountableAssumption}), the set $E$ can be selected
such that $\mu\left(\R_{\geq t_{0}}\setminus E\right)=0$. Since the
functions $\left(x,t\right)\to f_{\sigma}\left(x,t\right)$ are measurable,
\cite[Equation 27, p. 85]{Filippov1988} can be used to conclude that
$\forall\sigma\in\mathcal{N}$ there exist sets $\left\{ E_{\sigma}\subseteq\R_{\geq t_{0}}\right\} _{\sigma\in\mathcal{N}}$
with $\mu\left(\R_{\geq t_{0}}\setminus E_{\sigma}\right)=0,$ such
that for each $\sigma$ and $\forall\left(x,t\right)\in\R^{n}\times E_{\sigma}$,
there exists a measure zero set $N_{\sigma}\left(t\right)\subseteq\R^{n}$
such that $\mathbb{F}_{\sigma}\left(x,t\right)=\bigcap_{\delta>0}\overline{\co}\left\{ f_{\sigma}\left(y,t\right)\mid y\in\b\left(x,\delta\right)\setminus N_{\sigma}\left(t\right)\right\} $.
The selection $E=\cap_{\sigma\in\mathcal{N}}E_{\sigma}$ then satisfies
$\mu\left(\R_{\geq t_{0}}\setminus E\right)=0$, which, along with
(\ref{eq:Filippov1}) and (\ref{Filippov 2}), proves (\ref{eq:Convex Hull Almost Everywhere}). 
\end{IEEEproof}
\bigskip{}
\begin{IEEEproof}[Proof of Lemma \ref{lem:Reduction}]
The proof closely follows the proof of Lemma 1 in \cite{Bacciotti.Ceragioli1999}.
Let $x:\mathcal{I}\to\R^{n}$ be a solution of (\ref{eq:Inclusion})
such that $x\left(t_{0}\right)\in\mathcal{D}$. Consider the set of
times $\mathcal{T}\subseteq\left[t_{0},T\right)$ where $\dot{x}\left(t\right)$
is defined, $\dot{x}\left(t\right)\in F\left(x\left(t\right),t\right)$,
and $\dot{V}_{i}\left(x\left(t\right),t\right)$ is defined $\forall i\geq0$.
Since $x$ is a solution of (\ref{eq:Inclusion}) and the functions
$V_{i}$ are locally Lipschitz-continuous, $\mu\left(\left[t_{0},T\right)\setminus\mathcal{T}\right)=0$,
where $\mu$ denotes the Lebesgue measure on $\R$. The idea is to
show that $\dot{x}\left(t\right)\in\tilde{F}\left(x\left(t\right),t\right)$,
not just $F\left(x\left(t\right),t\right)$. Indeed since $V_{i}$
is locally Lipschitz-continuous, for $t\in\mathcal{T}$ its time derivative
can be expressed as 
\[
\dot{V}_{i}\left(x\left(t\right),t\right)=\lim_{h\to0}\frac{\left(V_{i}\left(x\left(t\right)+h\dot{x}\left(t\right),t+h\right)-V_{i}\left(x\left(t\right),t\right)\right)}{h}.
\]
Since each $V_{i}$ is regular, for $i\geq1$, $\dot{V}_{i}\left(x\left(t\right),t\right)=V_{i+}^{\prime}\left(\left[x\left(t\right);t\right],\left[\dot{x}\left(t\right);1\right]\right)=V_{i}^{o}\left(\left[x\left(t\right);t\right],\left[\dot{x}\left(t\right);1\right]\right)=\max\left(p^{T}\left[\dot{x}\left(t\right);1\right],p\in\partial V_{i}\left(x\left(t\right),t\right)\right),$
and $\dot{V}_{i}\left(x\left(t\right),t\right)=V_{i-}^{\prime}\left(\left[x\left(t\right);t\right],\left[\dot{x}\left(t\right);1\right]\right)=V_{i}^{o}\left(\left[x\left(t\right);t\right],\left[\dot{x}\left(t\right);1\right]\right)=\min\left(p^{T}\left[\dot{x}\left(t\right);1\right],p\in\partial V_{i}\left(x\left(t\right),t\right)\right),$
where $V_{+}^{\prime}$ and $V_{-}^{\prime}$ denote the right and
left directional derivatives and $V^{o}$ denotes the Clarke-generalized
derivative \cite[p. 39]{Clarke1990}. Hence, $p^{T}\left[\dot{x}\left(t\right);1\right]=\dot{V}_{i}\left(x\left(t\right),t\right),$
$\forall p\in\partial V_{i}\left(x\left(t\right),t\right)$, which
implies $\dot{x}\left(t\right)\in G_{i}\left(x\left(t\right),t\right)$
for each $i$. Therefore, $\dot{x}\left(t\right)\in\tilde{F}\left(x\left(t\right),t\right)$.
Hence, (\ref{eq:ReducedF}), along with the fact that $\dot{V}\left(x\left(t\right),t\right)=p^{T}\left[\dot{x}\left(t\right);1\right],$
$\forall p\in\partial V\left(x\left(t\right),t\right)$, implies that
$\forall t\in\mathcal{T}$, $\dot{V}\left(x\left(t\right),t\right)\leq-W\left(x\left(t\right)\right)$.
Since $\mu\left(\left[t_{0},T\right)\setminus\mathcal{T}\right)=0$,
$\dot{V}\left(x\left(t\right),t\right)\leq-W\left(x\left(t\right)\right)$
for almost all $t\in\left[t_{0},T\right)$.
\end{IEEEproof}
\textcolor{blue}{In the following, three technical Lemmas are stated
to facilitate the proof of Corollary \ref{cor:minmaxCts}.}
\begin{lem}
\textcolor{blue}{\label{lem:F-cts}If $\left\{ F_{\sigma}:\R^{n}\times\R_{\geq t_{0}}\rightrightarrows\R^{n}\mid\sigma\in\n\right\} $
is a collection of locally bounded, continuous, compact-valued, and
convex-valued maps, then the set-valued map $F\coloneqq\left(x,t\right)\mapsto\overline{\co}\bigcup_{\sigma\in\n}F_{\sigma}\left(x,t\right)$
is continuous.}
\end{lem}
\begin{IEEEproof}
\textcolor{blue}{Let $H:\R^{n}\times\R_{\geq t_{0}}\rightrightarrows\R^{n}$
be defined as $H\left(x,t\right)=\co\left(F_{1}\left(x,t\right)\cup F_{2}\left(x,t\right)\right)$.
If $N\subset\R^{n}$ is an open set containing $H\left(x,t\right)$,
then $\exists\epsilon>0$ such that $H\left(x,t\right)+\b\left(\left(x,t\right),\epsilon\right)\subset N$.
Since $F_{1}$ and $F_{2}$ are upper semicontinuous (USC), there
exist open sets $M_{1},M_{2}\subset\R^{n}\times\R_{\geq t_{0}}$ such
that $\left(x,t\right)\subset M_{1}\cap M_{2}$, $F_{1}\left(M_{1}\right)\subset H\left(x,t\right)+\b\left(\left(x,t\right),\epsilon\right)$,
and $F_{2}\left(M_{2}\right)\subset H\left(x,t\right)+\b\left(\left(x,t\right),\epsilon\right)$.
Therefore, $F_{1}\left(x,t\right)\cup F_{2}\left(x,t\right)\subset H\left(x,t\right)+\b\left(\left(x,t\right),\epsilon\right)$.
Since $H\left(x,t\right)+\b\left(\left(x,t\right),\epsilon\right)$
is convex, $\co\left(F_{1}\left(x,t\right)\cup F_{2}\left(x,t\right)\right)\subset H\left(x,t\right)+\b\left(\left(x,t\right),\epsilon\right)$.
Thus, $H$ is USC.}

\textcolor{blue}{It is easy to see that $\left(x,t\right)\mapsto F_{1}\left(x,t\right)\cup F_{2}\left(x,t\right)$
is lower semicontinuous (LSC). Using \cite[Theorem 5.9 (c)]{Rockafellar.Wets2009},
$H$ is also LSC. Inductively, the map $\left(x,t\right)\mapsto\co\cup_{k=1}^{K}F_{k}\left(x,t\right)$
is continuous $\forall K<\infty$. Thus, the collection $\left\{ F_{k}\right\} _{k\in\n}$
defined as $F_{k}\left(x,t\right)=\co\cup_{\sigma=1}^{k}F_{\sigma}\left(x,t\right)$
is a collection of nondecreasing continuous set-valued maps. By \cite[Exercise 4.3]{Rockafellar.Wets2009},
the sequence $\left\{ F_{k}\right\} _{k\in\n}$ converges pointwise
to the map $\left(x,t\right)\mapsto\overline{\cup_{k\in\n}F_{k}\left(x,t\right)}$.
Since the sets $\left\{ F_{k}\right\} $ are nested, $\overline{\cup_{k\in\n}F_{k}\left(x,t\right)}=\overline{\co\cup_{\sigma\in\n}F_{\sigma}\left(x,t\right)}$.
Hence, by \cite[Theorem 5.48 (a)]{Rockafellar.Wets2009}, the map
$\left(x,t\right)\mapsto\overline{\co}\cup_{\sigma\in\n}F_{\sigma}\left(x,t\right)$,
is continuous.}\footnote{\textcolor{blue}{By \cite[Theorem 5.7 (c)]{Rockafellar.Wets2009},
the notion of LSC in this paper is equivalent to the notion of inner
semicontinuity in \cite{Rockafellar.Wets2009}. Since the all the
maps under consideration are locally bounded and compact valued, by
\cite[Theorem 5.19]{Rockafellar.Wets2009}, the notion of USC in this
paper is equivalent to the notion of outer semicontinuity in \cite{Rockafellar.Wets2009}. }}\textcolor{blue}{{} }
\end{IEEEproof}
\begin{lem}
\textcolor{blue}{\label{lem:phi-cts}Let $g:\R^{n}\to\R$ be continuous
and let $F:\R^{n}\times\R_{\geq t_{0}}\rightrightarrows\R^{n}$ be
a locally bounded, continuous, and compact-valued map. If $\phi\coloneqq\left(x,t\right)\mapsto\max_{q\in F\left(x,t\right)}g\left(q\right)$,
then $\phi$ is continuous at $\left(x,t\right)$, $\forall\left(x,t\right)\in\R^{n}\times\R_{\geq t_{0}}$.}
\end{lem}
\begin{IEEEproof}
\textcolor{blue}{If not, then $\exists\epsilon>0$ such that $\forall\delta>0$,
$\exists\left(y,\tau\right)\in\b\left(\left(x,t\right),\delta\right)$
such that $\left|\phi\left(y,\tau\right)-\phi\left(x,t\right)\right|\geq\epsilon$.
If $\phi\left(y,\tau\right)-\phi\left(x,t\right)\geq\epsilon$ then
$\argmax_{q\in F\left(y,\tau\right)\cup F\left(x,t\right)}g\left(q\right)\subset F\left(y,\tau\right)\setminus F\left(x,t\right)$.
If $\phi\left(x,t\right)-\phi\left(y,\tau\right)\geq\epsilon$, then
$\argmax_{q\in F\left(y,\tau\right)\cup F\left(x,t\right)}g\left(q\right)\subset F\left(x,t\right)\setminus F\left(y,\tau\right)$.
That is, $\argmax_{q\in F\left(y,\tau\right)\cup F\left(x,t\right)}g\left(q\right)\subset F\left(x,t\right)\triangle F\left(y,\tau\right)$.
Let $\beta>0$. If $\left\{ \left(y_{k},\tau_{k}\right)\right\} _{k\in\n}\subset\overline{\b}\left(\left(x,t\right),\beta\right)$
is a sequence converging to $\left(x,t\right)$ such that $\left|\phi\left(y_{k},\tau_{k}\right)-\phi\left(x,t\right)\right|\geq\epsilon$,
then, $\forall k\in\n$, $\max_{q\in F\left(y_{k},\tau_{k}\right)\cup F\left(x,t\right)}g\left(q\right)=\max_{q\in F\left(x,t\right)\triangle F\left(y_{k},t_{k}\right)}g\left(q\right)$.
Since $g$ and $F$ are continuous and $F$ is locally bounded, the
sequence $\left\{ \max_{q\in F\left(y_{k},\tau_{k}\right)\cup F\left(x,t\right)}g\left(q\right)\right\} _{k\in\n}$
is a bounded sequence. On the other hand, since $F$ is continuous,
the sequence $\left\{ F\left(x,t\right)\triangle F\left(y_{k},\tau_{k}\right)\right\} _{k\in\n}$
converges to the null set, and hence, the sequence $\left\{ \max_{q\in F\left(y_{k},\tau_{k}\right)\cup F\left(x,t\right)}g\left(q\right)\right\} _{k\in\n}$
converges to $-\infty$, which is a contradiction.}
\end{IEEEproof}
\begin{lem}
\textcolor{blue}{\label{lem:h-cts}Let $g:\R^{n}\times\R^{n}\to\R$
be a continuous function and let $F:\R^{n}\times\R_{\geq t_{0}}\rightrightarrows\R^{n}$
be a locally bounded, USC, and compact-valued map. Let $h\coloneqq\left(p,x,t\right)\mapsto\max_{q\in F\left(x,t\right)}g\left(p,q\right)$.
If $C_{x}\subset\R^{n}\times\R_{\geq t_{0}}$ and $C_{p}\subset\R^{n}$
are compact, then $h$ is continuous in $p$, uniformly in $\left(x,t\right)$
over $C_{p}\times C_{x}$.}
\end{lem}
\begin{IEEEproof}
\textcolor{blue}{Since $g$ is continuous, and $F\left(C_{x}\right)$
and $C_{p}$ are compact,}\footnote{\textcolor{blue}{$F\left(C_{x}\right)$ is bounded by \cite[Lemma 15, p. 66]{Filippov1988},
and since $F$ is USC and $C_{x}$ is compact, $F\left(C_{x}\right)$
is also closed by \cite[Theorem 5.25 (a)]{Rockafellar.Wets2009}.}}\textcolor{blue}{{} it is uniformly continuous on $C_{p}\times F\left(C_{x}\right)$.
Thus, given $\epsilon>0$, $\exists\delta>0$, independent of $\left(p,x,t\right)$,
such that $\forall p,p_{0}\in C_{p}$ and $\forall q,q_{0}\in F\left(C_{x}\right)$,
$\left\Vert p-p_{0}\right\Vert <\delta\land\left\Vert q-q_{0}\right\Vert <\delta$
$\implies g\left(p_{0},p_{0}\right)<g\left(p,q\right)+\epsilon.$
In particular, $\left\Vert p-p_{0}\right\Vert <\delta\implies$ $g\left(p_{0},p_{0}\right)<g\left(p,q_{0}\right)+\epsilon.$
For any fixed $p_{0}\in C_{p}$ and $\left(x,t\right)\in C_{x}$,
$\exists q_{0}\in F\left(x,t\right)$ such that $h\left(p_{0},x,t\right)=g\left(p_{0},p_{0}\right)$,
and hence, $h\left(p_{0},x,t\right)<g\left(p,q_{0}\right)+\epsilon$.
Since $g\left(p,q_{0}\right)\leq h\left(p,x,t\right)$ by definition,
$h\left(p_{0},x,t\right)<h\left(p,x,t\right)+\epsilon$. That is,
$\forall p,p_{0}\in C_{p}$ and $\forall\left(x,t\right)\in C_{x}$,
$\left\Vert p-p_{0}\right\Vert <\delta\implies$ $h\left(p_{0},x,t\right)<h\left(p,x,t\right)+\epsilon.$
By symmetry, $\left|h\left(p_{0},x,t\right)-h\left(p,x,t\right)\right|<\epsilon$.}
\end{IEEEproof}
\textcolor{blue}{}
\begin{IEEEproof}[Proof of Corollary \textcolor{blue}{\ref{cor:minmaxCts}}]
\textcolor{blue}{Rademacher's theorem \cite[Theorem 3.2]{Evans.Gariepy2015}
and \cite[Proposition 2.3.6 (d)]{Clarke1990} imply that $\partial V$
is single-valued for almost all $\left(x,t\right)\in\R^{n}\times\R_{\geq t_{0}}$.
As a result, for almost all $\left(x,t\right)\in\R^{n}\times\R_{\geq t_{0}}$,
$\dot{\bar{V}}_{F}\left(x,t\right)=\underline{\dot{V}}_{F}\left(x,t\right)$.
By Proposition \ref{prop:Final}, for any $\left(x,t\right)\in\R^{n}\times\R_{\geq t_{0}}$
and $\beta>0$, there exists a sequence $\left\{ \left(y_{k},\tau_{k}\right)\right\} _{k\in\n}\subset\overline{\b}\left(\left(x,t\right),\beta\right)$,
converging to $\left(x,t\right)$ such that $\partial V\left(y_{k},\tau_{k}\right)=\left\{ \nabla V\left(y_{k},\tau_{k}\right)\right\} \eqqcolon\left\{ p_{k}\right\} $
and $\max_{q\in F\left(y_{k},\tau_{k}\right)}p_{k}^{T}\left[q;1\right]\leq-W\left(y_{k}\right).$}

\textcolor{blue}{Let $q_{k}\in\argmax_{q\in F\left(y_{k},\tau_{k}\right)}p_{k}^{T}\left[q;1\right].$
Since the set-valued map $F$ is locally bounded and USC, the sequence
$\left\{ q_{k}\right\} _{k\in\n}$ is bounded, and hence, admits a
convergent subsequence $\left\{ q_{k_{l}}\right\} _{l\in\n}$ converging
to some $q^{*}\in\R^{n}\times\R_{\geq t_{0}}$. Since $\partial V$
is locally bounded and USC (cf.\cite[p. 4]{Rockafellar1981}), the
sequence $\left\{ p_{k_{l}}\right\} _{l\in\n}$ is bounded. Hence,
there exists a subsequence $\left\{ p_{k_{l_{m}}}\right\} _{m\in\n}$
converging to some $p^{*}\in\R^{n}$. Hence, 
\begin{equation}
\left(p^{*}\right)^{T}\left[q^{*};1\right]\leq\lim_{m\to\infty}-W\left(y_{k_{l_{m}}}\right)=-W\left(x\right).\label{eq:WBound'}
\end{equation}
Using the characterization of the generalized gradient from \cite[p. 11, eq. (4)]{Clarke1990},
$p^{*}\in\partial V\left(x,t\right)$. From Lemma \ref{lem:F-cts},
$F$ is continuous, and hence, $q^{*}\in F\left(x,t\right)$.}

\textcolor{blue}{Let $h\coloneqq\left(p,x,t\right)\mapsto\max_{q\in F\left(x,t\right)}p^{T}\left[q;1\right]$.
To prove the corollary, it needs to be established that $h\left(p^{*},x,t\right)=\left(p^{*}\right)^{T}\left[q^{*};1\right]$.
The inequality $h\left(p^{*},x,t\right)\geq\left(p^{*}\right)^{T}\left[q^{*};1\right]$
is immediate from the definitions. Also, 
\begin{alignat}{1}
 & h\left(p^{*},x,t\right)-\left(p^{*}\right)^{T}\left[q^{*};1\right]=h\left(p^{*},x,t\right)-h\left(p^{*},y_{k_{l_{m}}},\tau_{k_{l_{m}}}\right)\nonumber \\
 & +h\left(p^{*},y_{k_{l_{m}}},\tau_{k_{l_{m}}}\right)-h\left(p_{k_{l_{m}}},y_{k_{l_{m}}},\tau_{k_{l_{m}}}\right)\nonumber \\
 & +h\left(p_{k_{l_{m}}},y_{k_{l_{m}}},\tau_{k_{l_{m}}}\right)-\left(p^{*}\right)^{T}\left[q^{*};1\right].\label{eq:epsilonover3}
\end{alignat}
Let $\varepsilon>0$. By definition of $p^{*}$ and $q^{*}$, $\exists M_{1}\in\n$
such that  $\forall m\geq M_{1}$, $\left|h\left(p_{k_{l_{m}}},y_{k_{l_{m}}},\tau_{k_{l_{m}}}\right)-\left(p^{*}\right)^{T}\left[q^{*};1\right]\right|<\frac{\varepsilon}{3}$.
Since $\partial V$ and $F$ are USC, $\partial V\left(\overline{\b}\left(\left(x,t\right),\beta\right)\right)$
and $F\left(\overline{\b}\left(\left(x,t\right),\beta\right)\right)$
are closed by \cite[Theorem 5.25]{Rockafellar.Wets2009}, and hence,
compact. Since $\left(p,q\right)\mapsto p^{T}\left[q;1\right]$ is
continuous, Lemma \ref{lem:h-cts} implies that the function $h$
is continuous in $p$, uniformly in $\left(x,t\right)$, over $\partial V\left(\overline{\b}\left(\left(x,t\right),\beta\right)\right)\times\overline{\b}\left(\left(x,t\right),\beta\right)$.
Hence, $\exists M_{2}\in\n$ such that $\forall m\geq M_{2}$, $\left|h\left(p^{*},y_{k_{l_{m}}},\tau_{k_{l_{m}}}\right)-h\left(p_{k_{l_{m}}},y_{k_{l_{m}}},\tau_{k_{l_{m}}}\right)\right|<\frac{\varepsilon}{3}.$
Lemma \ref{lem:phi-cts} implies that the function $\phi\coloneqq\left(x,t\right)\mapsto h\left(p^{*},x,t\right)$
is continuous. Hence, $\exists M_{3}>0$ such that $\forall m\geq M_{3}$,
$\left|h\left(p^{*},x,t\right)-h\left(p^{*},y_{k_{l_{m}}},\tau_{k_{l_{m}}}\right)\right|\leq\frac{\varepsilon}{3}$. }

\textcolor{blue}{Thus, for $m\geq\max\left\{ M_{1},M_{2},M_{3}\right\} $,
$h\left(p^{*},x,t\right)\leq\left(p^{*}\right)^{T}\left[q^{*};1\right]+\varepsilon$.
Since $\varepsilon$ was arbitrary, $h\left(p^{*},x,t\right)=\left(p^{*}\right)^{T}\left[q^{*};1\right]$.
Hence, from (\ref{eq:WBound'}) and the definition of $h$, $\exists p^{*}\in\partial V\left(x,t\right)$
such that $\max_{q\in F\left(x,t\right)}\left(p^{*}\right)^{T}\left[q;1\right]\leq W\left(x\right)$,
and hence, $\min_{p\in\partial V\left(x,t\right)}\max_{q\in F\left(x,t\right)}p^{T}\left[q;1\right]\leq-W\left(x\right).$}
\end{IEEEproof}
\begin{IEEEbiography}[{\includegraphics[clip,width=1in,height=1.25in]{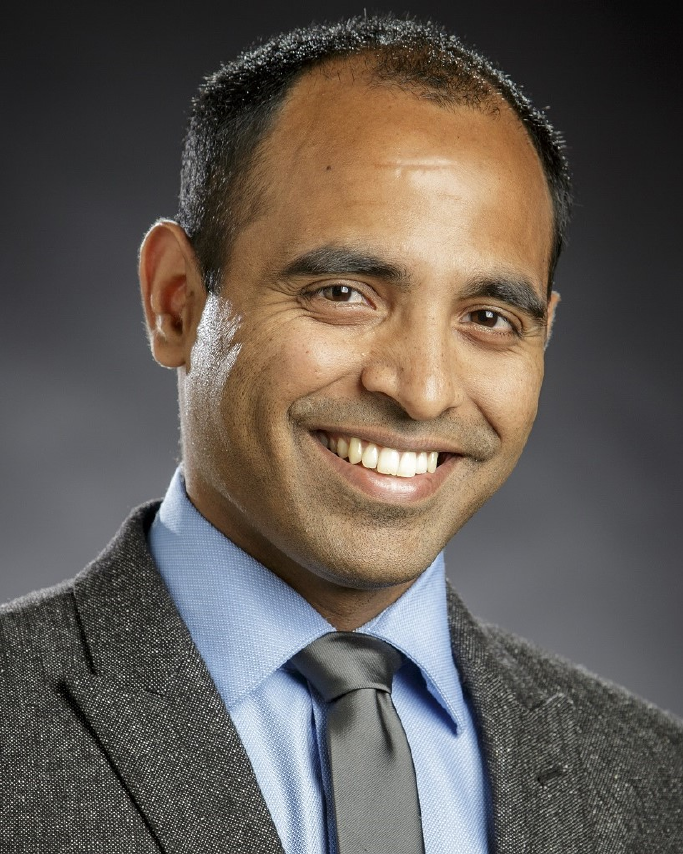}}]{Rushikesh Kamalapurkar}
 received his M.S. and Ph.D. degrees in 2011 and 2014, respectively,
from the Mechanical and Aerospace Engineering Department at the University
of Florida. After working for a year as a postdoctoral research fellow
with Dr. Warren E. Dixon, he was selected as the 2015-16 MAE postdoctoral
teaching fellow. In 2016 he joined the School of Mechanical and Aerospace
Engineering at the Oklahoma State University as an Assistant professor.
His primary research interest has been intelligent, learning-based
optimal control of uncertain nonlinear dynamical systems. He has published
3 book chapters, 18 peer reviewed journal papers and 19 peer reviewed
conference papers. His work has been recognized by the 2015 University
of Florida Department of Mechanical and Aerospace Engineering Best
Dissertation Award, and the 2014 University of Florida Department
of Mechanical and Aerospace Engineering Outstanding Graduate Research
Award.
\end{IEEEbiography}

\begin{IEEEbiography}[{\includegraphics[clip,width=1in,height=1.25in]{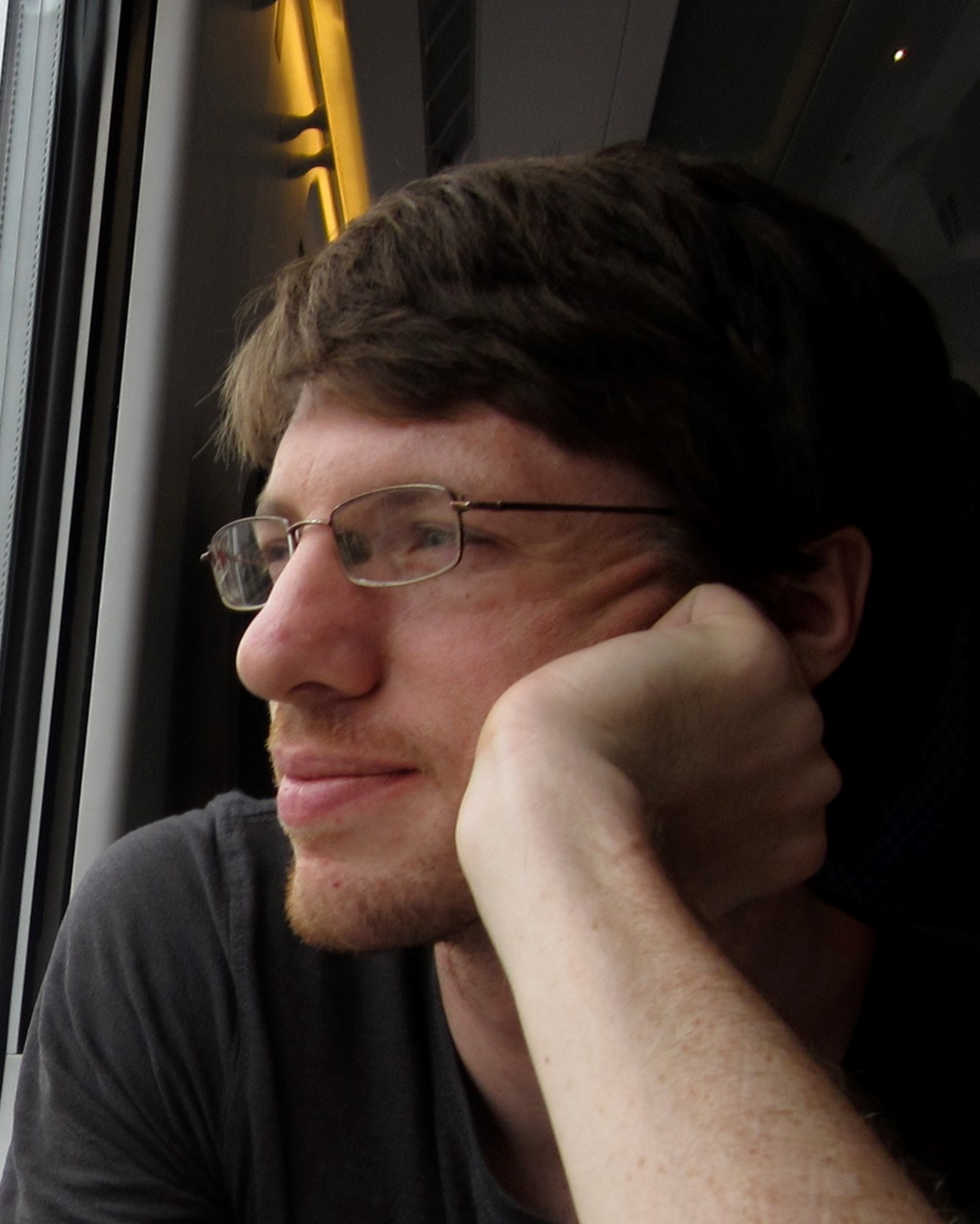}}]{Joel A. Rosenfeld}
 received his Ph.D. in Mathematics at the University of Florida
in 2013, under the advisement of Dr. Michael T. Jury. He joined the
department of Mechanical and Aerospace Engineering at the University
of Florida in 2013 as a postdoctoral researcher working with Dr. Warren
E. Dixon. His research focuses on approximation problems in control
theory.
\end{IEEEbiography}

\begin{IEEEbiography}[{\includegraphics[clip,width=1in,height=1.25in]{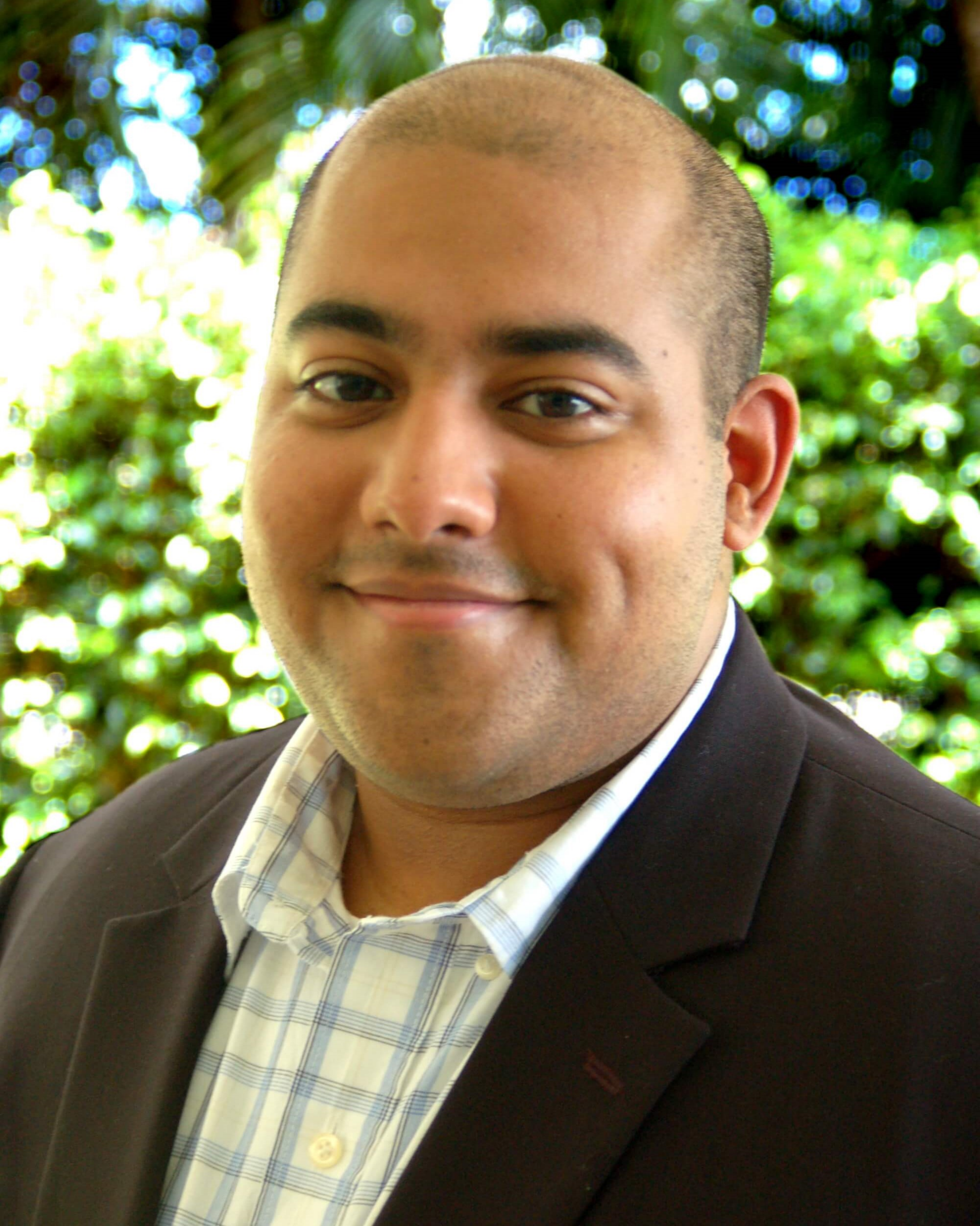}}]{Anup Parikh}
 received the B.S. degree in mechanical and aerospace engineering,
the M.S. degree in mechanical engineering, and the Ph.D. in aerospace
engineering from the University of Florida. Currently he is a Postdoctoral
Researcher at Sandia National Laboratories in Albuquerque, NM. His
primary research interests include Lyapunov-based control and estimation
theory and application in autonomous systems.
\end{IEEEbiography}

\begin{IEEEbiography}[{\includegraphics[clip,width=1in,height=1.25in]{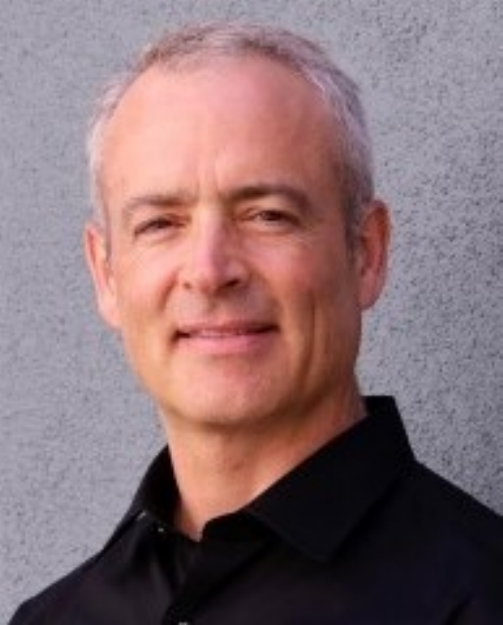}}]{Andrew R. Teel}
 received his A.B. degree in Engineering Sciences from Dartmouth
College in Hanover, New Hampshire, in 1987, and his M.S. and Ph.D.
degrees in Electrical Engineering from the University of California,
Berkeley, in 1989 and 1992, respectively. After receiving his Ph.D.,
he was a postdoctoral fellow at the Ecole des Mines de Paris in Fontainebleau,
France. In 1992 he joined the faculty of the Electrical Engineering
Department at the University of Minnesota, where he was an assistant
professor until 1997. Subsequently, he joined the faculty of the Electrical
and Computer Engineering Department at the University of California,
Santa Barbara, where he is currently a Distinguished Professor and
director of the Center for Control, Dynamical systems, and Computation.
His research interests are in nonlinear and hybrid dynamical systems,
with a focus on stability analysis and control design. He has received
NSF Research Initiation and CAREER Awards, the 1998 IEEE Leon K. Kirchmayer
Prize Paper Award, the 1998 George S. Axelby Outstanding Paper Award,
and was the recipient of the first SIAM Control and Systems Theory
Prize in 1998. He was the recipient of the 1999 Donald P. Eckman Award
and the 2001 O. Hugo Schuck Best Paper Award, both given by the American
Automatic Control Council, and also received the 2010 IEEE Control
Systems Magazine Outstanding Paper Award. In 2016, he received the
Certificate of Excellent Achievements from the IFAC Technical Committee
on Nonlinear Control Systems. He is Editor-in-Chief for Automatica,
and a Fellow of the IEEE and of IFAC.
\end{IEEEbiography}

\begin{IEEEbiography}[{\includegraphics[clip,width=1in,height=1.25in]{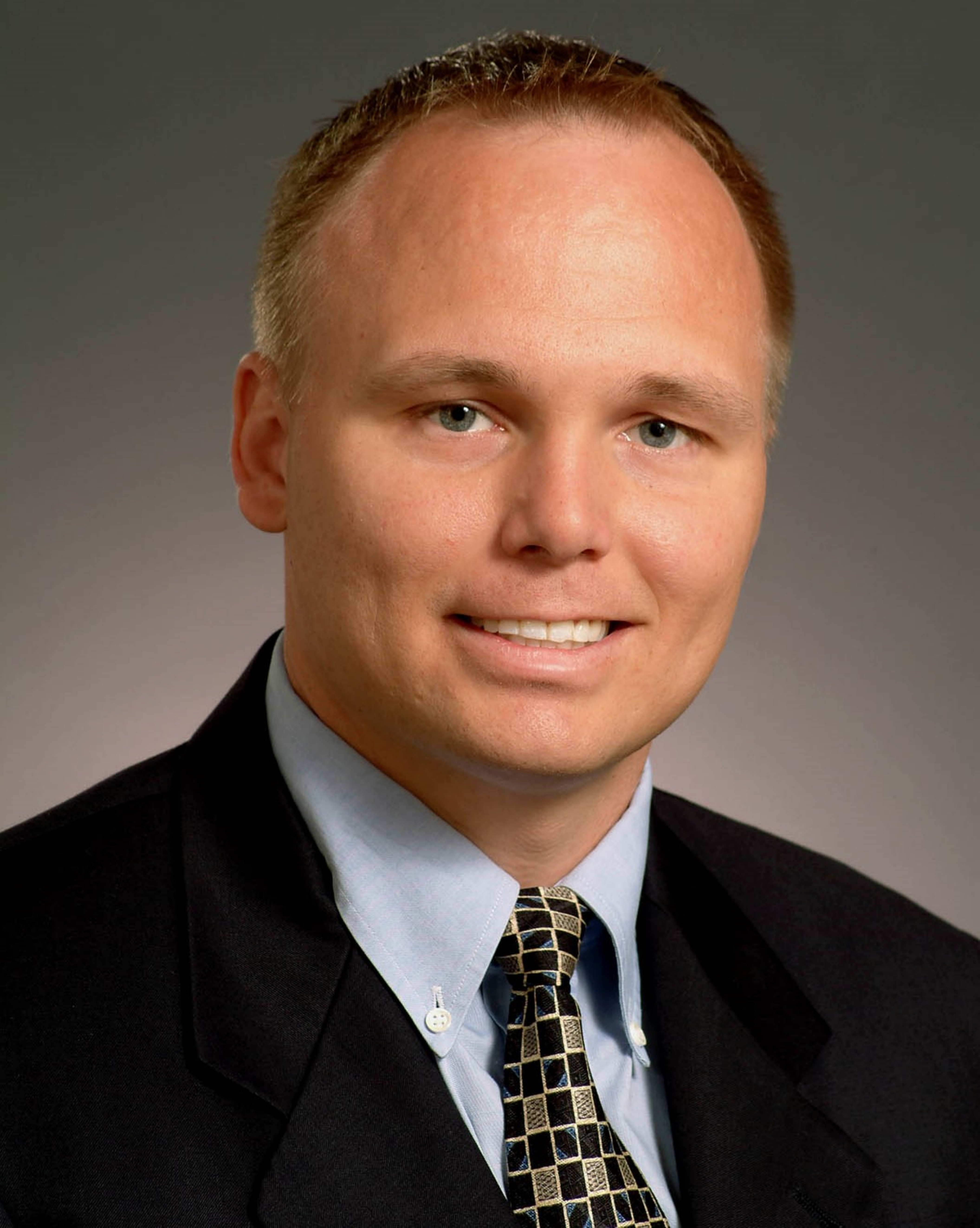}}]{Warren E. Dixon}
 received his Ph.D. in 2000 from the Department of Electrical and
Computer Engineering from Clemson University. He worked as a research
staff member and Eugene P. Wigner Fellow at Oak Ridge National Laboratory
(ORNL) until 2004, when he joined the University of Florida in the
Mechanical and Aerospace Engineering Department. His main research
interest has been the development and application of Lyapunov-based
control techniques for uncertain nonlinear systems. His work has been
recognized by the 2015 \& 2009 American Automatic Control Council
(AACC) O. Hugo Schuck (Best Paper) Award, the 2013 Fred Ellersick
Award for Best Overall MILCOM Paper, a 2012-2013 University of Florida
College of Engineering Doctoral Dissertation Mentoring Award, the
2011 American Society of Mechanical Engineers (ASME) Dynamics Systems
and Control Division Outstanding Young Investigator Award, the 2006
IEEE Robotics and Automation Society (RAS) Early Academic Career Award,
an NSF CAREER Award, the 2004 Department of Energy Outstanding Mentor
Award, and the 2001 ORNL Early Career Award for Engineering Achievement.
He is a Fellow of both ASME and IEEE, an IEEE Control Systems Society
(CSS) Distinguished Lecturer, and served as the Director of Operations
for the Executive Committee of the IEEE CSS Board of Governors (2012-2015).
He was awarded the Air Force Commander's Public Service Award (2016)
for his contributions to the U.S. Air Force Science Advisory Board.
He is currently or formerly an associate editor for ASME Journal of
Journal of Dynamic Systems, Measurement and Control, Automatica, IEEE
Transactions on Systems Man and Cybernetics: Part B Cybernetics, and
the International Journal of Robust and Nonlinear Control. 
\end{IEEEbiography}

\end{document}